# Nanoconfined, dynamic electrolyte gating and memory effects in multilayered graphene-based membranes

Jing Xiao[1,2*], Hualin Zhan[2*], Zaiquan Xu[1,3], Xiao Wang[2], Ke Zhang[1], Zhiyuan Xiong[2], George P. Simon[1], Zhe Liu[4], Dan Li[1,2 †]

[1]Department of Materials Science and Engineering, Monash University, VIC 3800, Australia

[2]Department of Chemical Engineering, The University of Melbourne, VIC 3010, Australia

[3]School of Mathematical and Physical Sciences, University of Technology Sydney, NSW, 2007, Australia

[4]Department of Mechanical Engineering, The University of Melbourne, VIC 3010, Australia

*These authors contributed equally to this work.

†Email: dan.li1@unimelb.edu.au

**Multilayered graphene-based nanoporous membranes with electrolyte incorporated between individual sheets is a unique nano-heterostructure system in which nanoconfined electrons in graphene and ions confined in between sheets are intimately coupled throughout the entire membrane. In contrast to the general notion that the electrolyte gating is unlikely to appear in multilayered graphene stacks, it is demonstrated in this work that the electrolyte gating effect in monolayer graphene can be transferred to its corresponding multilayered porous membranes. This gating effect presented on each individual graphene sheets through electrolyte confined in nanopores provides a real-time, electrical approach for probing the complex dynamics of nanoconfined electrical double layer. This has enabled the observation of the ionic memory effect in supercapacitors and produces new insights into the charging dynamics of supercapacitors. Such discoveries may stimulate the design of novel nanoionic devices.**

Ion transport in nanopores or nanochannels under dynamic conditions is ubiquitous in numerous applications including supercapacitors[1], capacitive deionization[2], artificial muscles[3] and nanofluidics[4]. Recent studies in nanoporous carbon-based supercapacitors have revealed that the dynamics of ions in nanoconfined electrical double layers (EDLs) can be very different from that in the bulk electrolyte due to the spatial confinement, ion-ion correlations, and enhanced interactions between ions and pore surfaces in nanopores[5-8]. Obtaining comprehensive understanding of nanoconfined ion dynamics not only provides valuable guidance for optimizing the existing applications of nanoporous materials[1,9], but can also stimulate the design of novel nanoionic devices. For example, a recent report shows that ion flow across a single nanopore can exhibit a hysteretic response to the applied potential (i.e., memory effect) at high frequency, making it possible to design new memory capacitors or memcapacitors from nanopores[10-12]. Recent simulations also predict that the phase behaviour of ions in porous electrodes with their pore sizes close to that of the ions could exhibit ferroelectric phase behaviours, implying new applications of nanoconfined ions beyond conventional energy storage[13].

An in-depth understanding of nanoconfined ion dynamics requires techniques for real-time probing of the ion movement in nanoporous materials but has proven to be challenging. Methods including nuclear magnetic resonance[7,14], electrochemical quartz crystal microbalance[15,16], infrared spectroscopy[17,18], and X-ray scattering[5,19] have all been used to probe the structure of EDL – a region of polarized ions at electrode-electrolyte interfaces that underpins many electrochemical applications. With these techniques, the information of the global change of ionic concentration and compositions of EDL inside supercapacitor electrodes can be obtained. However, ionic behaviour in EDL could evolve with time differently, depending on the structural, chemical, and electrical conditions, where the overall electrochemical performance of supercapacitors may be affected. Therefore, new *in situ* techniques are required to monitor the transient response of ions in EDL, where not only insights to their time-dependent behaviour can be obtained but also the contribution of ion dynamics to the electrochemical performance at the device level could be simultaneously evaluated in a facile manner.

We have recently demonstrated that large-area multilayered graphene-based membranes (MGMs) with their average interlayer distance tuneable from 10 nm down to subnanometre can be readily prepared by the controlled assembly of reduced graphene oxide (rGO) sheets[20-22]. When MGMs are immersed in an electrolyte where an external potential is applied, EDLs can immediately build up between monolayer graphene sheets, providing high EDL capacitance for energy storage[20]. From an electronic materials point of view, this membrane presents a unique nano-heterostructure of electrons and ions, in which electrons of graphene are restricted in the atomically thin sheets while ions are confined in the nanoscale gap between them. In this work, we demonstrate how the interactions of nano-confined electrons and ions in this membrane can lead to a new technique for electrically monitoring the ion dynamics of EDLs and how this technique assists in the discovery of a memcapacitive effect in nanoporous electrodes and provides new insights into the dynamics of supercapacitors.

**Hypothesis**

The existing techniques for probing the EDL dynamics are generally based on monitoring the ion concentration change within the nanoporous electrodes[9]. An alternative strategy to track the transient response of EDL is to measure in real-time the impact of EDL on the electrical property of the electrode material. It has been reported that when the electrode is made of a semiconductor such as silicon, the local electrical field across the EDL can affect the charge (electrons or holes) density in the semiconductive material, thereby modulating the electronic conductance of the electrode; hence the so-called electrochemical field effect or electrolyte gating effect[23,24]. The ion transport properties at the electrode interface could thus be indirectly monitored by measuring the conductance variation of the semiconductor, and to date has been used for developing chemical sensors[25,26]. To our knowledge, the electrolyte gating effect has been observed only at the interface of semiconductor and bulk electrolyte solutions[27]. We hypothesized that if the pore walls of a bulk nanoporous material are made of a semiconducting material and the electrolyte gating effect can be transferred to the bulk nanoporous material, the transient response of EDL could be monitored and read out electrically through the change of the electronic conductance of the porous materials.

Unlike graphite, monolayer graphene (primarily fabricated by mechanical cleavage or chemical vapour deposition) exhibits semiconductor-like properties because of the Fermi level is located near a relatively low electronic density of states[23,28]. However, an MGM contains thousands of chemically reduced graphene oxide (rGO), where the exact stacking order of individual monolayers is unclear, not to mention the presence of residual functional groups and defects may affect the electronic density of states. If the rGO sheets in the hydrogel MGM membranes are significantly separated from each other, and the contribution of defects to electronic density of states is not able to make rGO metallic[20,21,29], an electrolyte gating effect might be observable in MGM. Thus, the MGM material appears to be a good model system to verify the above hypothesis.

**Real-time monitoring the conductance of MGM during capacitive charging**

We adopted a conventional symmetric EDL-type capacitor design with two similar disks of MGMs serving as the working and counter (or gate) electrodes to investigate whether a potential applied to the working MGM electrode, which immediately leads to the formation of EDL within the electrode, would result in a change to its own electronic conductance ($G_{MGM}$). A platinum ring was placed on the top of the working MGM electrode to measure the $G_{MGM}$, where a low frequency (1 Hz) sinusoidal voltage with the amplitude of 1.0 mV is applied between the ring and current collector (Fig. 1a). This approach allowed the removal of the ionic conductance contribution to the measurement, and hence effectively decouples the electrical signal from the ion-electron network formed by the migration of ions in MGMs[30,31] (see details in Methods and Supplementary Fig. S1-S5).

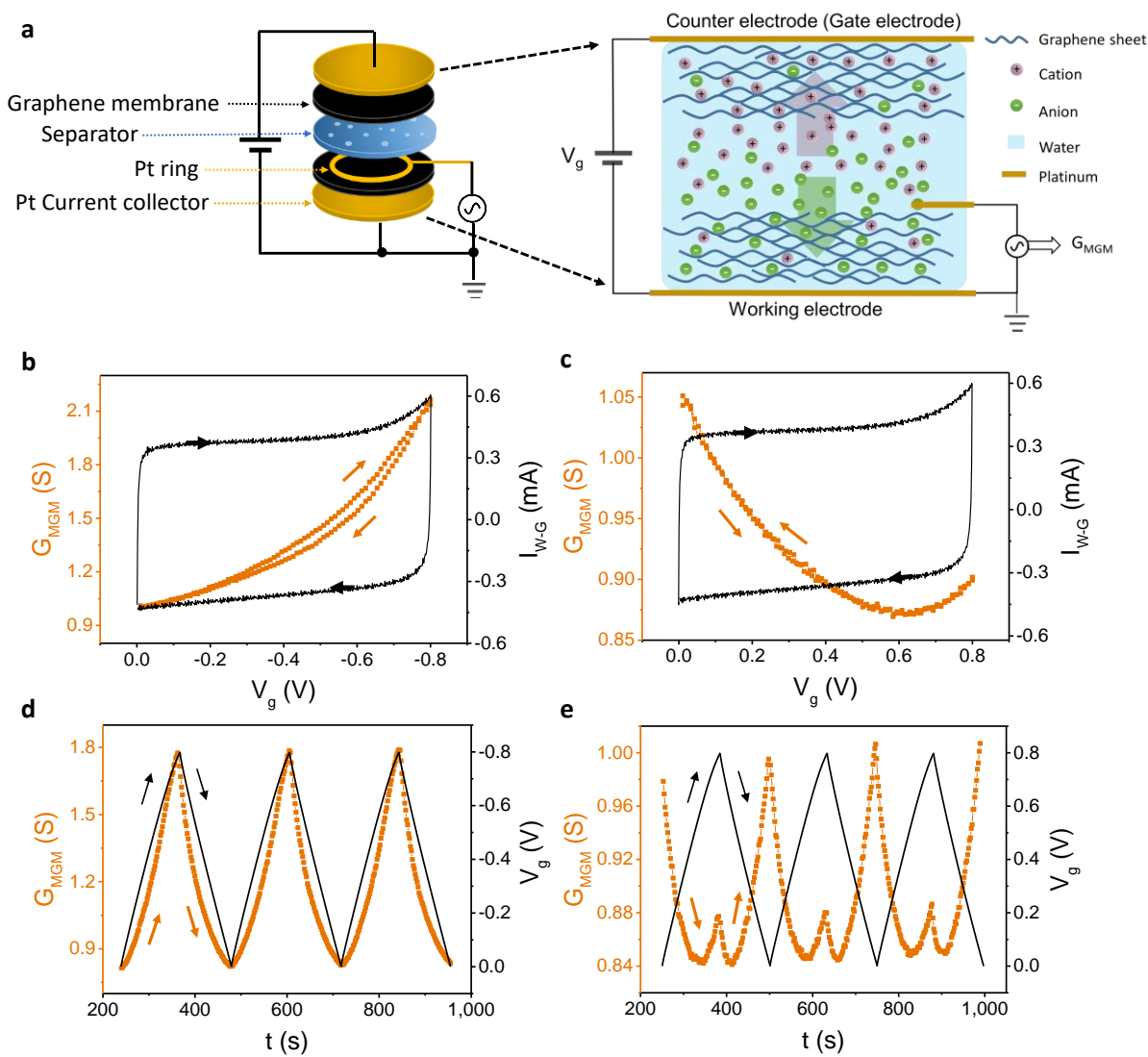

**Figure 1 | Characterization of an MGM-10 nm under electrolyte gating in 1.0 M KCl**. **a**, Schematic of the setup used for characterizing the electrolyte gating effect of MGM. The gate voltage ($V_g$) is defined as the voltage applied on the counter (or gate) electrode with respect to the working electrode. **b,c**, Conductance response of MGMs ($G_{MGM}$, trans-membrane conductance) to cyclic voltammetry scan under the voltage window of (b) -0.8~0 V and (c) 0~0.8 V, respectively. The scan rate is 5 mV s$^{-1}$. **d,e**, Conductance response of MGMs to galvanostatic charging-discharging under the voltage window of (d) -0.8~0 V and (e) 0~0.8 V, respectively. The charging current is 0.5 A g$^{-1}$. The interlayer spacing of the MGM is ~10 nm, with a diameter of 1.27 cm and thickness of ~100 μm.

We first monitor the $G_{MGM}$ *in operando* by charging the MGM with an interlayer distance of ~10 nm (MGM-10 nm) in 1.0 M KCl using the classic two-electrode cyclic voltammetry (CV) technique. The rectangular CV loops (black curves) shown in Fig. 1b and c indicate a typical capacitive behaviour. Note that our conventional three-electrode electrochemical measurements confirm that redox reactions are not involved in the charging process (Supplementary Fig. S4). When the voltage applied on the counter MGM ($V_g$, with respect to the working MGM electrode) changes from 0 to -0.8 V (Fig. 1b), the measured $G_{MGM}$ increases monotonically from 1.01 S to 2.17 S.

When $V_g$ is changed from 0 to 0.8 V (Fig. 1c), the resultant CV curve is similar to that of the negative charging in terms of both the shape and area (or the capacitance of the MGM, which is around 142.3 F g$^{-1}$ at a scan rate of 5 mV s$^{-1}$). However, the variation of $G_{MGM}$ during the positive charging displays a very different pattern. It decreases first from ~1.05 S at 0 V to ~0.87 S at 0.6 V, and then rises slightly afterwards, varying in a non-monotonic manner.

We further characterize the response of $G_{MGM}$ to $V_g$ by charging MGMs at a constant current using the method of galvanostatic cycling potential limitation technique (GCPL), which is widely used in characterization of electrochemical performance of energy storage devices. Again, the $G_{MGM}$ increases continuously with charging time when $V_g$ is negative (Fig. 1d) and shows a non-monotonic variation when the working MGM is positively charged (Fig. 1e). Similar potential-dependent $G_{MGM}$ and its response to the sign of $V_g$ are also observed in other electrolytes such as aqueous solutions of potassium trifluoromethanesulfonimide (KTFSI), 1-ethyl-3-methylimidazolium tetrafluoroborate ($EMIMBF_4$) and 1-butyl-3-methylimidazolium tetrafluoroborate ($BMIMBF_4$) (see Supplementary Fig. S6).

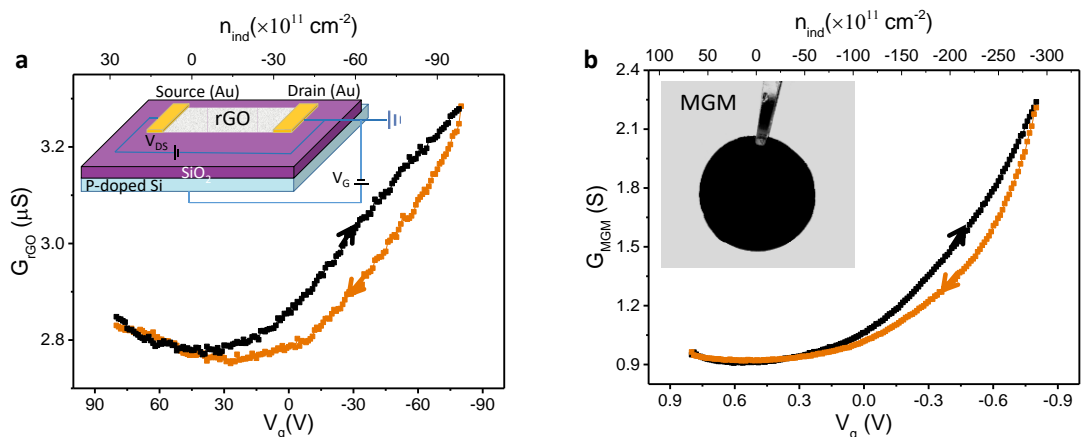

**Figure 2 | Comparison of the electrolyte gating effect of MGM with the electric field effect of a single rGO sheet under back gating. a**, on-sheet conductance of rGO ($G_{rGO}$) as a function of gate voltage ($V_g$) under back-gating. A small voltage of 10 mV ($V_{sd}$) was applied between the source and drain electrodes. The thickness of $SiO_2$ is around 285 nm. $n_{ind}$ is the induced charge carrier density in rGO. Inset: the schematic diagram of the experimental setup. **b**, $G_{MGM}$ of MGM-10 nm as a function of $V_g$ in 1.0 M KCl. The scanning rate is 5 mV s$^{-1}$. The inset is a photograph of the MGM-10 nm sample.

### Electrolyte-gating effect in MGM and potential for ion dynamics study

Our analysis reveals that the potential dependence of $G_{MGM}$ is unlikely to be caused by the possible expansion or shrinkage of MGM or the MGM/Pt contact during the capacitive charging (see detailed discussion in Supplementary Fig. S7). Rather, it can be explained by the electric field or electrolyte gating effect of the rGO sheets, where the charge carrier density in rGO is modulated by a local electrical field from the EDL. Previous studies have revealed that differing from an ideal pristine graphene sheet whose Dirac point is located at $V_g=0$[32], rGO behaves like a *p*-type semiconductor due to the existence of oxygen-containing groups and defects, and its Dirac point is shifted away from $V_g=0$[33]. To further confirm the semiconductive nature of our rGO, we fabricated a back-gated field effect transistor based on a single layer of rGO sheet as shown in Fig. 2a (see the characterization in Supplementary Fig. S8). The measured conductance was found to vary with the applied voltage, suggesting that the charge carrier density (according to the x-axis on the top of the figure) is modulated by $V_g$. This gating effect was further confirmed by the observation of minimum values of conductance when $V_g$ is around 30 V, as a result of the lowest intrinsic electron density near the "Dirac" point[23,34]

To clearly demonstrate how the $G_{MGM}$ is modulated by the external potential applied in our electrochemical setup, the relationship of $G_{MGM}$ against $V_g$ under both positive and negative polarisations is combined and plotted in Fig. 2b. The overall trend of $G_{MGM}$-$V_g$ looks very similar to that in Fig. 2a and the minimum of $G_{MGM}$ appears when $V_g>0$. In analogy to Fig. 2a where the oxide layer separates graphene and gate, EDL forms at the interface between each graphene sheet and the electrolyte where the gate electrode is immersed, *i.e.,* electrolyte gating. Similar to Fig. 2a, the occurrence of minimum of $G_{MGM}$ during the positive polarisation can be ascribed to the shift of "Dirac" point of rGO sheets as a result of the chemical doping and the electrostatic interaction between electrolyte and rGO described by the Gerischer model[35] (see Supplementary Fig. S15 for more detailed discussion). As a result of the shift of "Dirac" point to a positive voltage, the $G_{MGM}$-$V_g$ relation under negative polarisation is monotonic (Fig. 1b) while that under positive polarisation is non-monotonic (Fig. 1c). When $V_g$ is away from the "Dirac" point under negative polarisation, the number of majority charge carriers in graphene should be proportional to the number of ions in EDL. Therefore, the monotonic $G_{MGM}$-$V_g$ relationship suggests that the number of charge carriers in graphene changes, and hence the number of ions stored in EDL, with the applied voltage. This indicates that measurement of $G_{MGM}$ under negative polarisation could be used to study the charging/discharging process in EDL.

We have also calculated the density of states in MGMs under the electrolyte gating, which is estimated to vary from $7.5 \times 10^{12}$ to $-2.8 \times 10^{13}$ cm$^{-2}$ between 0.8 and -0.8 V, as opposed to that from $3 \times 10^{12}$ to $-1 \times 10^{13}$ cm$^{-2}$ between 80 and -80 V in conventional back gating (see detailed calculation and discussion in the Supplementary document). The significantly smaller value of $V_g$ in the electrolyte gating is because the EDL capacitance is several orders of magnitude greater than the capacitance in the back gating where SiO$_2$ is used as dielectric[24]. This in turn significantly increases the conductance of rGO in the electrolyte gating setup, while $V_g$ only changes slightly. Therefore, the technique of electrolyte gating is well suited for probing the ionic behaviour in EDL, as the low operation voltage prevents undesirable redox reactions.

The successful observation of electrolyte gating effect in MGM demonstrates an effective technique to electrically modulate charge carriers in an ensemble of graphene sheets, regardless of the exact stacking order. The large number of layers appear not to prevent the observation of electrolyte gating effect. This suggests that ions in electrolyte can access individual rGO sheets. Since the electrolyte gating effect on bulk graphite is not observable, as the very small inter-graphene distance (~ 0.34 nm) prevents insertions of ions, Fig. 2b also confirms that the individual rGO sheets in MGM remain largely separated.

The finding that the electrolyte gating effect is transferrable to the multilayered rGO network suggests that the MGM membrane, when charged in an electrolyte, could be used to provide a unique ion/electron nanohybrid system to understand the dynamics of ions confined in between individual sheets through a collective gating effect. The electrical conductivity of a solid material is proportional to the product of the mobility and density of charge carriers (electrons or holes in the case of graphene). According to the Gauss' law, the net number of counter ions stored in EDL and on the surface of the rGO sheets in MGM (denoted as $Q_{EDL}$) is the same as the net number of charges in rGO sheets when they are charged in an electrolyte. Provided that the electron/hole mobility of rGO is nearly independent of the gate potential when the potential applied is away from the Dirac point[36,37], $G_{MGM}$ should be proportional to $Q_{EDL}$. Since the response of $G_{MGM}$ to the variation of $Q_{EDL}$ is nearly instantaneous (less than microseconds as a result of the high mobility of rGO[38]), any change in $Q_{EDL}$ would be immediately reflected by the variation of $G_{MGM}$. Thus, the differential capacitance of MGM ($C_d$), which directly reflects the transient change of $Q_{EDL}$ with respect to voltage, can also be directly monitored as $C_d = dQ_{EDL}/dV \propto dG_{MGM}/dV$. Hence, the measurement of $G_{MGM}$ can be used to *in situ* monitor the transient state of EDL within the porous MGM membrane in real time.

**Probing nano-confined ion dynamics and memory effect of capacitance**

The MGM used in the above experiments has a relatively large average interlayer distance of ~10 nm[29]. To examine whether the interlayer distance ($d$), or the level of nanoconfinement, has any effect on the electrolyte gating effect, we prepare a series of MGMs with the interlayer distance varied from 5.0 nm, 2.0 nm, 0.8 nm, to 0.6 nm (denoted as MGM-5.0 nm, MGM-2.0 nm, MGM-0.8 nm, MGM-0.6 nm, respectively, see details in Supplementary Materials and Methods) and perform the electrolyte gating experiments in different electrolytes (see Supplementary Fig. S9 and Fig. S10). The electrolyte gating effect is observed in all cases examined, though the ratio of the gating effect is reduced when the interlayer distance is reduced. Of particular interest is that new phenomena appear when the interlayer distance is reduced below 1.0 nm, particularly when an electrolyte of a larger ionic size is used. Below we present the typical results obtained from the 1.0 M aqueous solution of BMIMBF$_4$ where the bare equivalent spherical diameters of the cation and anion are reported to be 0.68 and 0.46 nm, respectively[39,40]. As shown in Fig. 3a and b, when MGM-5.0 nm or MGM-2.0 nm is used as the working electrode, the variation of $G_{MGM}$ with $V_g$ during a CV scan (5 mV s$^{-1}$) shows a similar trend as that of MGM-10 nm. However, when MGM-0.8 nm or MGM-0.6 nm is employed (Fig. 3c and d), $G_{MGM}$ displays a remarkable hysteresis during the reverse scan. This hysteresis phenomenon is also observed in other electrolytes such as EMIMBF$_4$ and KTFSI and under both negative and positive polarisations (Supplementary Figs. S14 and S15).

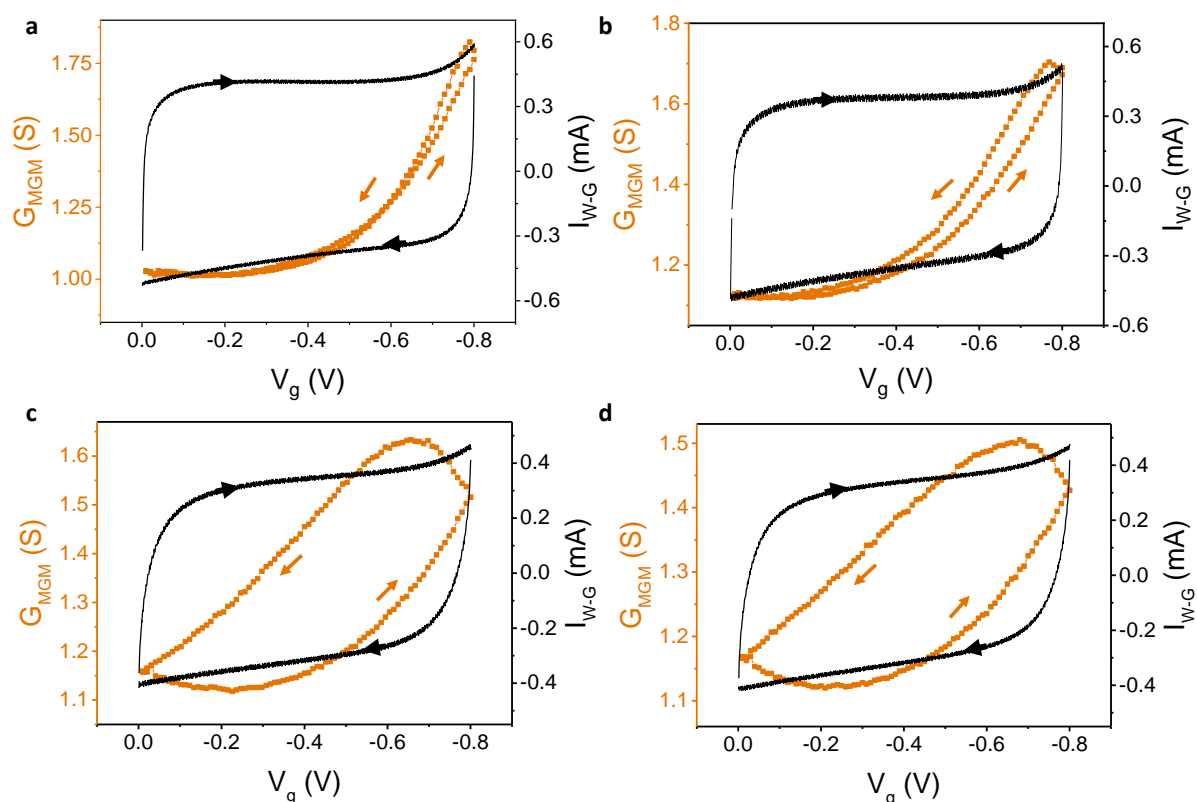

**Figure 3 | Nanoconfined electrolyte gating of MGMs under negative polarisation.** $G_{MGM}$ shows an obvious hysteresis with regard to $V_g$ when the interlayer distance of MGM is below 1.0 nm. $G_{MGM}$ and $I_{W-G}$ as a function of $V_g$ are shown for MGM with an interlayer distance of **a**, 5.0 nm, **b**, 2.0 nm, **c**, 0.8 nm, and **d**, 0.6 nm. The scan rate is 5 mV s$^{-1}$ and the electrolyte is 1.0 M BMIMBF$_4$. The results under positive polarisation are presented in Figure S14 and S15.

The GCPL experiments also reveal similar nanoconfinement-dependent hysteresis. As shown in Fig.4 a and b, when the charging current is 0.5 A g$^{-1}$, the $G_{MGM}$ of MGM-5.0 nm and MGM-2.0 nm varies almost synchronously with $V_g$, showing negligible hysteresis. In contrast, the $G_{MGM}$ of MGM-0.8 nm and MGM-0.6 nm reaches the peak value a few seconds after $V_g$ has attained the maximum, revealing a delay. The dependence of this delay on the interlayer distance of MGMs is analysed by measuring the time lag between peak $G_{MGM}$ and peak $V_g$. The calculated time lag is 4.0 s and 3.9 s for MGM-5.0 nm and MGM-2.0 nm, respectively, while it increases pronouncedly to 8.4 s and 15.7 s for MGM-0.8 nm and MGM-0.6 nm. Note that for clear comparison, only the results obtained from negative gating under which $G_{MGM}$ follows a monotonic linear dependence on the gate voltage are presented in Fig. 4. The delay of the peak $G_{MGM}$ to the peak voltage is also found during the positive gating, but in a more complicated fashion (see an example in Supplementary Fig. S11).

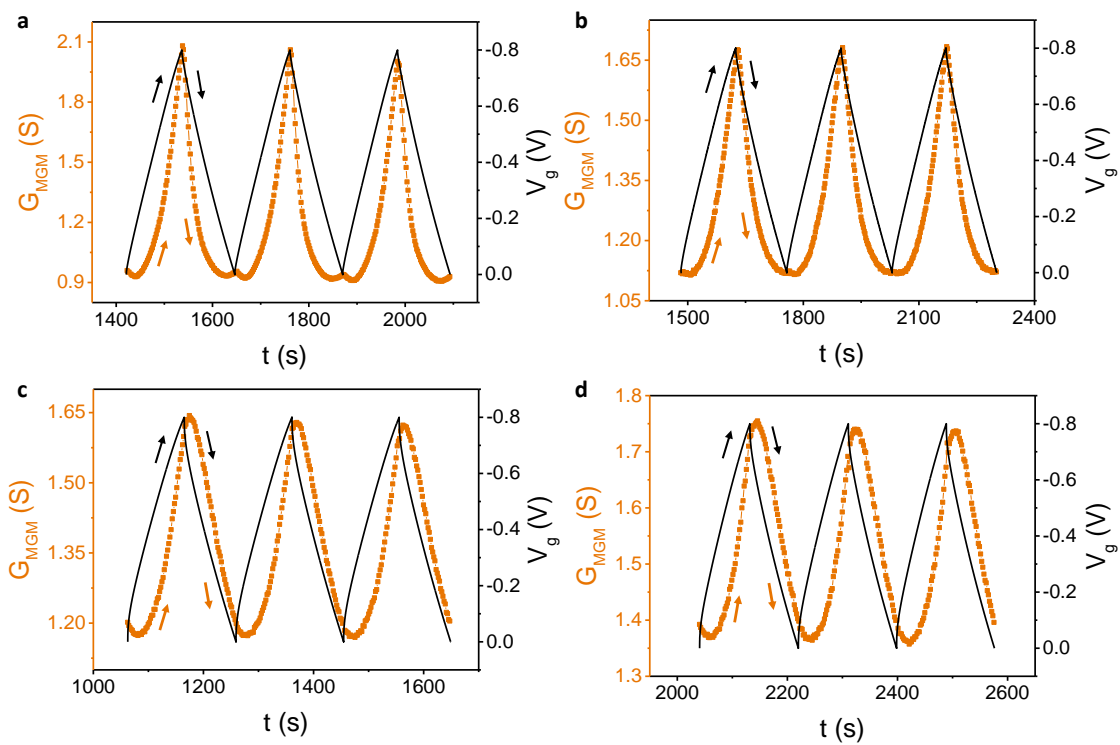

**Figure 4 | Characterization of the hysteretic response of $G_{MGM}$ under nanoconfinement using the GCPL technique.** Under this condition, $I_{W-G}$ varies in phase with $V_g$ and charges supplied by external circuit attains the maximum at the peak of $V_g$[41,42]. $G_{MGM}$ and $I_{W-G}$ as a function of $V_g$ are shown for MGM with an interlayer distance of **a**, 5.0 nm, **b**, 2.0 nm, **c**, 0.8 nm, and **d**, 0.6 nm. The charging rate is 0.5 A g$^{-1}$ and the electrolyte is 1.0 M BMIMBF$_4$.

As $G_{MGM}$ can be used to monitor the transient charges stored within the MGM, the time delay in nonlinear response of the $G_{MGM}$ to voltage under the extreme nanoconfinement indicates that the capacitance of the MGM in this case is history-dependent and exhibits a memory effect. To the best of our knowledge, this is the first direct experimental observation of memcapacitance effect in supercapacitors based on bulk nanoporous electrodes.

The memory effect can be further evidenced by the variation of $G_{MGM}$ under the open-circuit voltage (OCV) test. As shown in Fig. 5a, for MGM-5.0 nm, while $V_g$ starts to decrease instantly after the charging is switched off (black dashed line) due to its self-discharge, $G_{MGM}$ follows the same trend immediately (orange dashed line). In contrast, for MGM-0.8 nm, while $V_g$ displays a similar decrease as that of MGM-5.0 nm during the OCV measurement, $G_{MGM}$ continues to rise in the first 72 seconds after the circuit is opened. This means that the differential capacitance of MGM ($C_d$) must be negative in this case as $C_d = dQ_{EDL}/dV \propto dG_{MGM}/dV$. We note that negative differential capacitance has previously been observed in some ferroelectric ceramics under non-equilibrium conditions[43,44], which is attributed to the energy barrier formed between two degenerate polarization states. The experimental observation of the negative differential capacitance in our experiments suggests that ions confined in nanopores could also exhibit different states, rendering ferroelectric-like properties under certain conditions.

We have further tested other electrolytes including KTFSI which shares the same cation with KCl and EMIMBF$_4$ which has the same anion with BMIMBF$_4$ (see Supplementary Fig. S14 and S15). The $G_{MGM}$ is responsive to the gating voltage in all cases tested and the overall trend is similar to the results of KCl and BMIMBF$_4$. Remarkable hysteresis also appears in the cases of KTFSI and EMIMBF$_4$ for the MGM-0.8 nm sample, but it is insignificant for the MGM-5.0 nm sample. We also notice that the $G_{MGM}$-$V_g$ curve is more or less varied among different electrolytes and depends on the type of polarisation, suggesting that the nature of electrolytes play a secondary role in the electrolyte gating effect. This coincides with the finding that the capacitance of a nanoporous electrode is related to the nature of electrolytes, particularly when the ionic size of the electrolyte is relatively larger than the pore size of electrodes[8]. These results indicate that the electrolyte gating method could potentially be developed as a very sensitive ion probing technique if its relationship with the chemistry of electrolytes can be calibrated in the future. This also means that in principle, it should be possible to purposely design the molecular structure of electrolytes to realize memcapacitors with desired hysteresis or memory behaviors in the future.

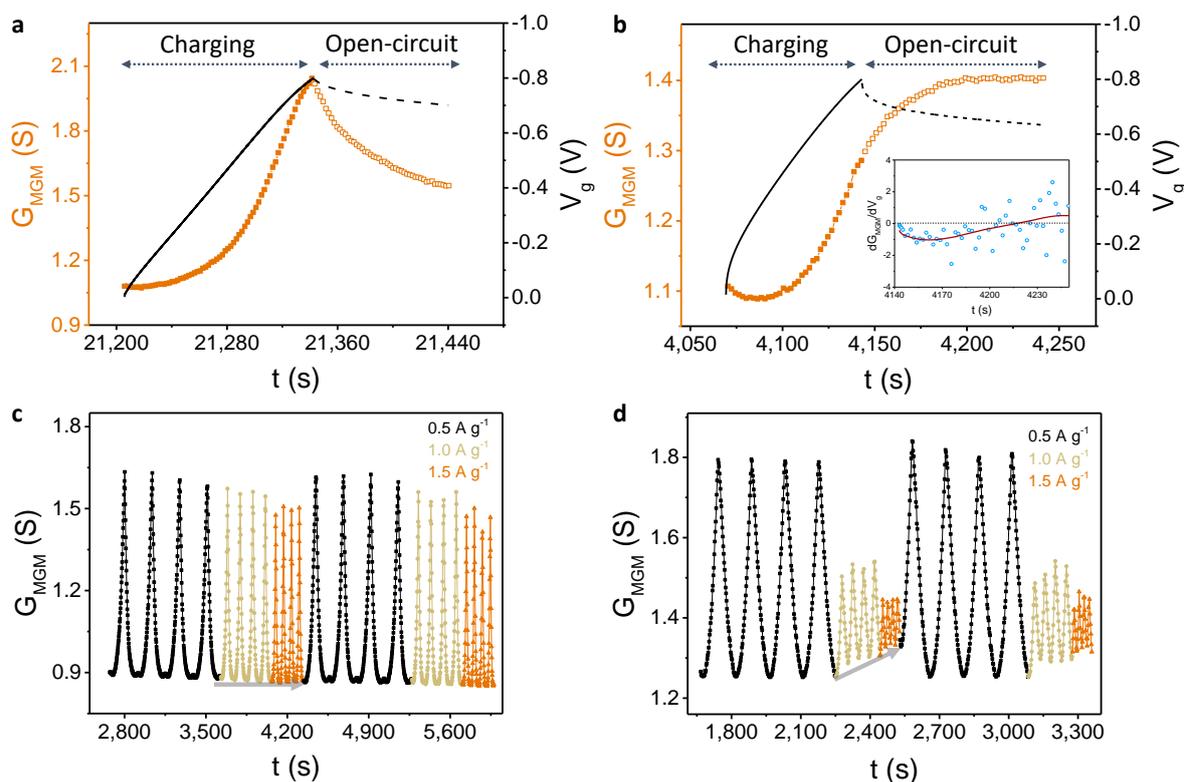

**Figure 5 | Characterization of memcapacitive effect (time-dependent $G_{MGM}$) during open-circuit process.** $G_{MGM}$ as a function of time is shown during the charging and open-circuit process for **a**, MGM-5.0 nm and **b**, MGM -0.8 nm. The inset in (b) shows the time-dependent $dG_{MGM}/dV_g$ (proportional to $C_d$) only during the open-circuit measurement. The charging current is 0.5 A g$^{-1}$ and the electrolyte is 1.0 M BMIMBF$_4$. **c-d,** The variation of $G_{MGM}$ for MGM-5.0 nm (c) and MGM-0.8 nm (d) at various charging rates in 1.0 M BMIMBF$_4$.

Recent *in situ* experiments and computational modellings have shown that the charging mechanism of supercapacitors is complex and can involve counter-ion adsorption, ion exchange, and/or co-ion desorption[9]. How the nano-confined EDL evolves during the charging/discharging depends on a broad range of parameters such as the electrode and electrolyte materials used, the size and surface chemistry of pores, the polarization of the electrode, the operation rate[45], as well as the ion concentration, the cation-to-anion size ratio, solvation energies[9,14,18,46-48], ionophobicity[6], and ion-ion correlations[49]. Such a great complexity makes it difficult to fully understand the exact mechanism of the nanoconfinement-induced memory effect at the stage. Plausible mechanisms could include slow ionic polarization due to the intrinsic finite mobility of ions, where the ionic current retains its original sign after the external voltage has reversed the direction. Increasing the scan rate of the voltage results in a larger hysteresis loop of $G_{MGM}$ for the samples of MGM-0.8 nm and MGM-0.6 nm (Supplementary Fig. S12), indicating that slow ionic polarization does play a role in the delayed response. On the other hand, in the GCPL measurements, the peak value of $G_{MGM}$ occurs a few seconds after the ionic current has reversed its sign (Fig.4a, d), implying that other factors may also be at play. Recent studies have shown that ions can reorganize at the electrolyte/electrode interface, which displays an ultraslow relaxation with a time scale of ~ 10 s$^{50,51}$. It is speculated that such reorganization or steric effect could be either enhanced or restricted under nanoconfinement and contributes to the memcapacitive effect. It would be of great interest to combine the electrolyte gating method with other techniques including multiscale modelling to advance the understanding of nano-confined ion dynamics in the future.

**Potential implications of nano-confined electrolyte gating effect**

The discovery of the electrolyte gating effect in the nanoporous membrane may have significant implications both practically and fundamentally. This technique provides a new *in operando* electrical method to monitor the dynamics of electrical double layer and is expected to be useful to gain new fundamental insights in dynamic electrochemical applications. For example, it has been observed that supercapacitors, particularly those made of extremely small pores, suffer a significant capacitance degradation at increased charging rates, which has been previously ascribed to the loss of accessible surface area[52]. These electrolyte gating results suggest that the memory effect of capacitance could also contribute. As shown in Fig. 5c and d, MGM-5.0 nm and MGM-0.8 nm are charged at various rates and the variation of $G_{MGM}$ are simultaneously monitored. The peak value of $G_{MGM}$ for MGM-5.0 nm and MGM-0.8 nm shows both a decreasing trend with charging rates, which is in line with the decrease in the $Q_{EDL}$ as well as $C_{EDL}$ (integral capacitance). When the current density changed from 0.5 A g$^{-1}$ to 1.5 A g$^{-1}$, the $C_{EDL}$ decreased from 148.1 F g$^{-1}$ to 141.3 F g$^{-1}$ for MGM-5.0 nm and from 95.1 F g$^{-1}$ to 48.5 F g$^{-1}$ for MGM-0.8 nm. However, for MGM-0.8 nm, a notable rise of minimum value of $G_{MGM}$ with charging rates is also observed. In addition, a similar increase in the lowest value of $G_{MGM}$ is also found in various electrolytes (Supplementary Fig. S13). Considering that the minimum value of $G_{MGM}$ corresponds to the $Q_{EDL}$ at the discharged state of EDL under a dynamic charging process, this result clearly indicates that more charges are retained in the nanopores at increased operation rates during the discharging process and are inactive for practical power delivery. It should be the memory effect-induced detention of charges rather than the loss of accessible surface area that has led to the deterioration of capacitance at increased operation rates.

Supercapacitors based on nanoporous electrodes have been overwhelmingly exploited for capacitive energy storage in the past. Our discoveries may point to some new unconventional applications for them. As discussed above, the memory effect and negative differential capacitance observed in our experiments suggests that ions restricted in extremely small pores could exhibit different states and could display a ferroelectric-like behaviour under dynamic conditions. The finding that these properties are dependent on the chemical nature of electrolytes, their interactions with graphene as well as the level of nanoconfinement implies that the memory effect could be potentially tailored to achieve desired phase behaviours through fine tuning of the pore nanostructure and electrolyte chemistry. We note that memory devices including memcapacitors are being considered as candidates for future neuromorphic computing devices[10,53]. Given that the electrolyte gating effect provides a simple electrical approach to transcribe ionic memory behaviours to an external circuit, it would be of great interest to further understand the mechanisms of the electrolyte gating effects in electroactive nanopores and examine their potential for future design of novel ionotronic devices[54] such as the neuromorphic information processors[55].

Although only reduced graphene oxide is used to demonstrate the key concept in this work, our preliminary results have shown that the electrolyte gating effect also exists in other porous materials comprised of single-wall carbon nanotubes and other 2D nanomaterials such as $MoS_2$. There appears much opportunity to exploit the interactions between nanoconfined electrons and ions in mixed conductors for both fundamental and applied applications.

**References**

1    Salanne, M. *et al.* Efficient storage mechanisms for building better supercapacitors. *Nat. Energy* **1**, 16070 (2016).
2    Werber, J. R., Osuji, C. O. & Elimelech, M. Materials for next-generation desalination and water purification membranes. *Nat. Rev. Mater.* **1**, 16018 (2016).
3    Foroughi, J. *et al.* Torsional carbon nanotube artificial muscles. *Science* **334**, 494-497 (2011).

4   Koltonow, A. R. & Huang, J. Two-dimensional nanofluidics. *Science* **351**, 1395-1396 (2016).
5   Futamura, R. *et al.* Partial breaking of the Coulombic ordering of ionic liquids confined in carbon nanopores. *Nat. Mater.* **16**, 1225 (2017).
6   Kondrat, S., Wu, P., Qiao, R. & Kornyshev, A. A. Accelerating charging dynamics in subnanometre pores. *Nat. Mater.* **13**, 387-393 (2014).
7   Forse, A. C. *et al.* Direct observation of ion dynamics in supercapacitor electrodes using in situ diffusion NMR spectroscopy. *Nat. Energy* **2**, 16216 (2017).
8   Chmiola, J. *et al.* Anomalous increase in carbon capacitance at pore sizes less than 1 nanometer. *Science* **313**, 1760-1763 (2006).
9   Forse, A. C., Merlet, C., Griffin, J. M. & Grey, C. P. New Perspectives on the Charging Mechanisms of Supercapacitors. *J. Am. Chem. Soc.* **18**, 5731-5744 (2016).
10  Krems, M., Pershin, Y. V. & Di Ventra, M. Ionic memcapacitive effects in nanopores. *Nano Lett.* **10**, 2674-2678 (2010).
11  Wang, D. *et al.* Transmembrane potential across single conical nanopores and resulting memristive and memcapacitive ion transport. *J. Am. Chem. Soc.* **134**, 3651-3654 (2012).
12  Dias, C., Ventura, J. & Aguiar, P. in *Advances in Memristors, Memristive Devices and Systems*   (eds Sundarapandian Vaidyanathan & Christos Volos)  305-342 (Springer International Publishing, 2017).
13  Kiyohara, K., Soneda, Y. & Asaka, K. Ferroelectric phase behaviors in porous electrodes. *Langmuir* **33**, 11574-11581 (2017).
14  Griffin, J. M. *et al.* In situ NMR and electrochemical quartz crystal microbalance techniques reveal the structure of the electrical double layer in supercapacitors. *Nat. Mater.* **14**, 812–819 (2015).
15  Levi, M. D. *et al.* Electrochemical quartz crystal microbalance (EQCM) studies of ions and solvents insertion into highly porous activated carbons. *J. Am. Chem. Soc.* **132**, 13220-13222 (2010).
16  Tsai, W.-Y., Taberna, P.-L. & Simon, P. Electrochemical Quartz Crystal Microbalance (EQCM) Study of Ion Dynamics in Nanoporous Carbons. *J. Am. Chem. Soc.* **24**, 8722-8728 (2014).
17  Richey, F. W., Dyatkin, B., Gogotsi, Y. & Elabd, Y. A. Ion Dynamics in Porous Carbon Electrodes in Supercapacitors Using in Situ Infrared Spectroelectrochemistry. *J. Am. Chem. Soc.* **135**, 12818-12826 (2013).
18  Richey, F. W., Tran, C., Kalra, V. & Elabd, Y. A. Ionic Liquid Dynamics in Nanoporous Carbon Nanofibers in Supercapacitors Measured with in Operando Infrared Spectroelectrochernistry. *J. Phys. Chem. C* **118**, 21846-21855 (2014).
19  Prehal, C. *et al.* Quantification of ion confinement and desolvation in nanoporous carbon supercapacitors with modelling and in situ X-ray scattering. *Nat. Energy* **2**, 16215 (2017).
20  Yang, X., Cheng, C., Wang, Y., Qiu, L. & Li, D. Liquid-mediated dense integration of graphene materials for compact capacitive energy storage. *Science* **341**, 534-537 (2013).
21  Yang, X. *et al.* Ordered gelation of chemically converted graphene for next‐generation electroconductive hydrogel films. *Angew. Chem. Int. Edit.* **50**, 7325-7328 (2011).
22  Li, D., Mueller, M. B., Gilje, S., Kaner, R. B. & Wallace, G. G. Processable aqueous dispersions of graphene nanosheets. *Nat. Nanotechnol.* **3**, 101-105 (2008).
23  Novoselov, K. S. *et al.* Electric field effect in atomically thin carbon films. *Science* **306**, 666-669 (2004).
24  Das, A. *et al.* Monitoring dopants by Raman scattering in an electrochemically top-gated graphene transistor. *Nat. Nanotechnol.* **3**, 210-215 (2008).
25  Heller, I. *et al.* Influence of electrolyte composition on liquid-gated carbon nanotube and graphene transistors. *J. Am. Chem. Soc.* **132**, 17149-17156 (2010).
26  Ohno, Y., Maehashi, K., Yamashiro, Y. & Matsumoto, K. Electrolyte-gated graphene field-effect transistors for detecting pH and protein adsorption. *Nano Lett.* **9**, 3318-3322 (2009).
27  Chhowalla, M., Jena, D. & Zhang, H. Two-dimensional semiconductors for transistors. *Nat. Rev. Mater.* **1**, 16052 (2016).
28  Eda, G., Fanchini, G. & Chhowalla, M. Large-area ultrathin films of reduced graphene oxide as a transparent and flexible electronic material. *Nat. Nanotechnol.* **3**, 270-274 (2008).
29  Cheng, C. *et al.* Ion transport in complex layered graphene-based membranes with tuneable interlayer spacing. *Sci. Adv.* **2**, e1501272 (2016).
30  Albery, W. J., Elliott, C. M. & Mount, A. R. A transmission line model for modified electrodes and thin layer cells. *J. Electroanal. Chem.* **288**, 15-34 (1990).
31  Saab, A. P., Garzon, F. H. & Zawodzinski, T. A. Determination of ionic and electronic resistivities in carbon/polyelectrolyte fuel-cell composite electrodes. *J. Electrochem. Soc.* **149**, A1541-A1546 (2002).

32  Schedin, F. *et al.* Detection of individual gas molecules adsorbed on graphene. *Nat. Mater.* **6**, 652-655 (2007).
33  Eda, G. & Chhowalla, M. Chemically Derived Graphene Oxide: Towards Large-Area Thin-Film Electronics and Optoelectronics. *Adv. Mater.* **22**, 2392-2415 (2010).
34  He, Q. *et al.* Transparent, flexible, all-reduced graphene oxide thin film transistors. *ACS Nano* **5**, 5038-5044 (2011).
35  Memming, R. Semiconductor Electrochemistry. *Weinheim, Germany: WILEY-VCH* (2015).
36  Zhan, H., Cervenka, J., Prawer, S. & Garrett, D. J. Molecular detection by liquid gated Hall effect measurements of graphene. *Nanoscale* **10**, 930-935 (2018).
37  Kim, H., Kim, B. J., Sun, Q., Kang, M. S. & Cho, J. H. Graphene Transistors Gated by Salted Proton Conductor. *Adv. Electron. Mater.* **2**, 1600122 (2016).
38  Wang, S. *et al.* High mobility, printable, and solution-processed graphene electronics. *Nano Lett.* **10**, 92-98 (2009).
39  Murgia, S., Monduzzi, M., Lopez, F. & Palazzo, G. Mesoscopic structure in mixtures of water and 1-butyl-3-methyl imidazolium tetrafluoborate: a multinuclear NMR study. *J. Solution Chem.* **42**, 1111-1122 (2013).
40  Jin, H. *et al.* Physical properties of ionic liquids consisting of the 1-butyl-3-methylimidazolium cation with various anions and the bis(trifluoromethylsulfonyl)imide anion with various cations. *J. Phys. Chem. B* **112**, 81-92 (2008).
41  Wen, C. J., Boukamp, B., Huggins, R. & Weppner, W. Thermodynamic and mass transport properties of "LiAl". *J. Electrochem. Soc.* **126**, 2258-2266 (1979).
42  Weppner, W. & Huggins, R. A. Determination of the kinetic parameters of mixed‐conducting electrodes and application to the system Li3Sb. *J. Electrochem. Soc.* **124**, 1569-1578 (1977).
43  Catalan, G., Jiménez, D. & Gruverman, A. Ferroelectrics: Negative capacitance detected. *Nat. Mater.* **14**, 137-139 (2015).
44  Khan, A. I. *et al.* Negative capacitance in a ferroelectric capacitor. *Nat. Mater.* **14**, 182-186 (2015).
45  He, Y. D., Huang, J. S., Sumpter, B. G., Kornyshev, A. A. & Qiao, R. Dynamic Charge Storage in Ionic Liquids-Filled Nanopores: Insight from a Computational Cyclic Voltammetry Study. *J. Phys. Chem. Lett.* **6**, 22-30 (2015).
46  Griffin, J. M. *et al.* Ion counting in supercapacitor electrodes using NMR spectroscopy. *Faraday Discuss.* **176**, 49-68 (2014).
47  Dyatkin, B. *et al.* Influence of Surface Oxidation on Ion Dynamics and Capacitance in Porous and Nonporous Carbon Electrodes. *J. Phys. Chem. C* **120**, 8730-8741 (2016).
48  Prehal, C. *et al.* Tracking the structural arrangement of ions in carbon supercapacitor nanopores using in situ small-angle X-ray scattering. *Energ. Environ. Sci.* **8**, 1725-1735 (2015).
49  Cheng, C., Jiang, G. P., Simon, G. P., Liu, J. Z. & Li, D. Low-voltage electrostatic modulation of ion diffusion through layered graphene-based nanoporous membranes. *Nat. Nanotechnol.* **13**, 685 (2018).
50  Uysal, A. *et al.* Structural origins of potential dependent hysteresis at the electrified graphene/ionic liquid interface. *J. Phys. Chem. C* **118**, 569-574 (2013).
51  Nishi, N., Hirano, Y., Motokawa, T. & Kakiuchi, T. Ultraslow relaxation of the structure at the ionic liquid| gold electrode interface to a potential step probed by electrochemical surface plasmon resonance measurements: asymmetry of the relaxation time to the potential-step direction. *Phys. Chem. Chem. Phys.* **15**, 11615-11619 (2013).
52  Conway, B. & Pell, W. Power limitations of supercapacitor operation associated with resistance and capacitance distribution in porous electrode devices. *J. Power Sources* **105**, 169-181 (2002).
53  Vaidyanathan, S. & Volos, C. *Advances in memristors, memristive devices and systems*. Vol. 701 (Springer, 2017).
54  Yang, C. & Suo, Z. Hydrogel ionotronics. *Nat. Rev. Mater.* **3**, 125-142 (2018).
55  Chun, H. & Chung, T. D. Iontronics. *Annu. Rev. Anal. Chem.* **8**, 441-462 (2015).

# Acknowledgements

We would like to acknowledge the financial support from the Australia Research Council and the University of Melbourne. This work made use of the facilities at the Monash Centre for Electron Microscopy (MCEM) and Melbourne Centre for Nanofabrication (MCN).

# Competing interests

The authors declare no competing interests.

# Data availability

The data that support the findings of this study are available from the corresponding author upon reasonable request.

**Additional information**
**1. Materials and methods**

**Fabrication of MGM and the tuning of its interlayer distance**
**Measuring the electrolyte gating effect of MGM**
**Exploiting the AC method to measure the electronic conductance of MGM** (Fig. S1- S3)
**Investigation into the influence of faradaic reaction and Ohmic contact on the conductance measured** (Fig. S4, S5)
**Examining the nature of electrolyte gating effect**
**2. Supplementary Figures** (Fig. S6- S15)

**3. Supplementary References**

# Supplementary information

# Nanoconfined, dynamic electrolyte gating and memory effects in multilayered graphene-based membranes

Jing Xiao[1,2*], Hualin Zhan[2*], Zaiquan Xu[1, 3], Xiao Wang[2], Ke Zhang[1], Zhiyuan Xiong[2], George P Simon[1], Zhe Liu[4], Dan Li[1,2†]

[1] Department of Materials Science and Engineering, Monash University, VIC 3800, Australia.

[2] Department of Chemical Engineering, the University of Melbourne, VIC 3010, Australia

[3] School of Mathematical and Physical Sciences, University of Technology Sydney, NSW, 2007, Australia

[4]Department of Mechanical Engineering, The University of Melbourne, VIC 3010, Australia

[*]These authors contributed equally to this work.

[†]Correspondence to:  dan.li1@unimelb.edu.au.

**Content**

**1. Materials and methods**

    **Fabrication of MGM and the tuning of its interlayer distance**

    **Measuring the electrolyte gating effect of MGM**

    **Exploiting the AC method to measure the electronic conductance of MGM** (Fig. S1-S3)

    **Investigation into the influence of faradaic reaction and Ohmic contact on the conductance measured** (Fig. S4, S5)

    **Examining the nature of electrolyte gating effect**

**2. Supplementary Figures** (Fig. S6- S15)

**3. Supplementary References**

## 1. Materials and Methods

**Fabrication of MGMs and fine-tuning of their interlayer distance**

Reduced graphene oxide (rGO) dispersion was prepared using the method previously reported[1]. Briefly, graphene oxide aqueous solution (0.5 mg mL$^{-1}$, 100 mL), 0.2 mL hydrazine (35 wt% in water) and 0.35 mL ammonia solution (28 wt% in water) were mixed thoroughly in a glass vial, and then the vial was placed in a water bath (~100 °C) for 3 hours to reduce the graphene oxide.

Multilayered graphene hydrogel membranes (MGMs) were prepared using the method we reported previously[2]. Specifically, 30 mL of the as-prepared rGO solution was vacuum-filtrated via an Isopore™ polycarbonate membrane (Merck Millipore Ltd, 0.1 μm pore size). The vacuum (50 mPa) was immediately disconnected when no free solution remained. The MGM was then peeled off from the filter membrane and soaked in deionized water overnight to remove the residual ammonia and unreacted hydrazine.

Interlayer spacing of the membranes was tuned using the capillary compression method we reported previously[3]. The MGM was first thoroughly exchanged with $H_2SO_4$ aqueous solution of controlled concentration. Water inside the gel membranes was then selectively removed through vacuum evaporation, during which capillary compression force between rGO layers resulted in the uniform compression of membrane along the thickness direction. With $H_2SO_4$ retained in the gel membrane, the average interlayer spacing was readily tuned by controlling the concentration of $H_2SO_4$ aqueous solution. Subsequently, the as-prepared membranes were washed thoroughly with deionized water to remove the $H_2SO_4$ inside the membrane porous structure and immersed in the testing electrolytes for 3 hours to fully exchange with electrolytes before testing. In this work, the concentrations of the $H_2SO_4$ aqueous solution were 5.0 M, 1.0 M and 0.1 M to make MGMs with interlayer distances of around 5.0 nm, 2.0 nm and 0.8 nm, respectively. The average interlayer spacing in MGM was estimated from the packing density of MGM using the method previously reported[3-5]. The thickness of the MGM membranes was measured via scanning electron microscopy (FEI Nova Nano SEM450).

**Measuring the electrolyte gating effect of MGM**

Electrolyte gating experiments were conducted with a customised T-cell (Fig.1 a) in which two MGMs with a diameter of 12.7 mm were attached onto two separate platinum electrodes. An electrically insulating glass fibre membrane (ADVANTEC, GA-55) was placed in between the MGMs to avoid short circuit while ions in electrolyte were still allowed to travel through freely. The device is similar to a conventional supercapacitor design. Since individual rGO sheets in the MGM remain largely separated with a certain interlayer spacing[3], this setup enables ions to travel to the rGO sheets across the whole MGM upon the application of a voltage to form EDL on each individual rGO sheets of the MGM (Fig. 1b). To reflect the electrolyte gating nature, this voltage is referred to as gate voltage in this article. As reported in the main text, the formation of EDL on the rGO surface would cause its electronic conductance to vary[6-8], which is also known as the electrolyte gating effect. The electronic conductance of MGM as a function of gate voltage was measured to examine the electrolyte gating effect in the MGM using the following procedure.

As shown in Fig. 1a, a platinum ring was inserted between one of the MGMs and the glass fibre membrane to measure the electronic conductance of MGM. The impedance of MGM was obtained by applying a sinusoidal alternating voltage (amplitude: 1.0 mV, frequency: 1 Hz) between this ring and the platinum electrode to which the MGM was attached (Fig. 1a). Previously, a DC voltage is generally applied between the drain and source electrodes and the metal/graphene contacts are sealed with electrically insulating resins to avoid the direct contact of metal electrodes to the electrolyte solution, and therefore to decouple the ionic contribution to the conductance of graphene films [7,9-11]. However, in our case, the MGM membranes are full of electrolyte and can indeed be viewed as a type of mixed electronic-ionic conductor. Given that the metal (Pt in our case) would inevitably be in contact with the electrolyte solution, the ionic conductance couldn't be decoupled if a DC current is applied, which will complicate the analysis of the electrolyte gating effect. Previous studies have shown that the electronic conductance of a mixed electronic-ionic conductor can be individually obtained by applying an AC voltage with low frequencies[12-16], with which the influence of the DC current between the working and gate electrode can be largely separated. Detailed discussions on the validity of our method to measure the electronic conductance of MGM ($G_{MGM}$) and its role in examining the electrolyte gating effect are shown in the following sections.

**Exploiting the AC measurement method to measure the electronic conductance of MGM**

In order to confirm that the AC measurement method adapted from Ref. 15 for measuring the electronic conductance of MGM is reliable, we first perform an equivalent circuit model analysis and then validate it with experiments.

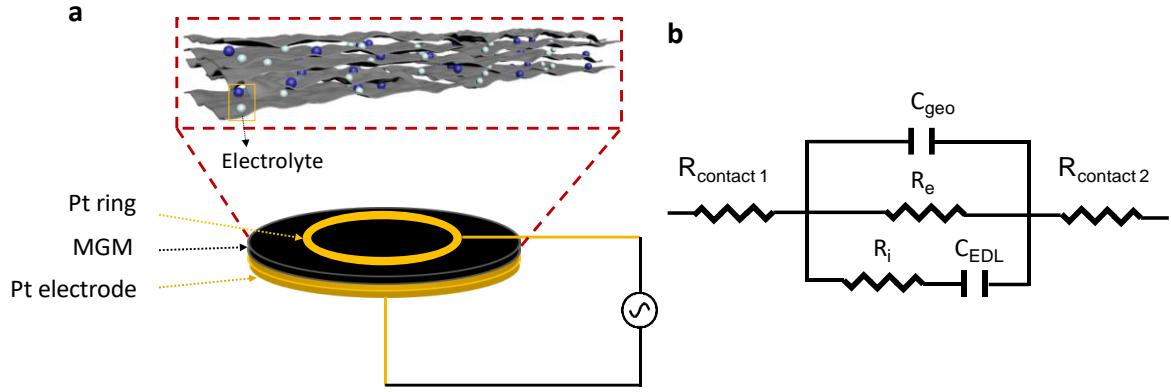

**Figure S1. Equivalent circuit model analysis of the AC measurement of G$_{MGM}$.** Schematic of the experimental setup for measuring the electronic conductance of MGM (**a**) and equivalent circuit of the testing system (**b**). R$_{contact\ 1}$ is the contact resistance between the Pt ring and MGM; R$_{contact\ 2}$ is the contact resistance between Pt electrode (current collector) and MGM; C$_{geo}$ corresponds to the geometric capacitance contributed by Pt ring and Pt electrode; R$_e$ is the electronic conductance of the MGM and R$_i$ is ionic resistance of electrolyte inside the MGM; C$_{EDL}$ is the EDL capacitance of MGM.

The total impedance (Z) of the circuit shown in Fig. S1 is

$$Z = R_{contact\ 1} + R_{contact\ 2} + \frac{1}{\frac{1}{R_e} + i\omega C_{geo} + Y_i} \tag{S1}$$

Where $Y_i = \frac{1}{R_i + \frac{1}{i\omega C_{EDL}}}$ is the admittance of the electrolyte and $\omega$ is the frequency of the alternating sinusoidal voltage. R$_{contact\ 1}$ is the contact resistance between the Pt ring and MGM; R$_{contact\ 2}$ is the contact resistance between Pt current collector and MGM; C$_{geo}$ corresponds to the geometric capacitance contributed by Pt ring and Pt electrode; R$_e$ is the electronic conductance of the MGM and R$_i$ is ionic resistance of electrolyte inside the MGM; C$_{EDL}$ is the EDL capacitance of MGM. Note that electron transfer resistance within EDL can be neglected here, as the amplitude of the sinusoidal voltage (1 mV) is unlikely to induce a significant redox reaction. If the frequency is sufficiently low, $Y_i \approx i\omega C_{EDL}$. Therefore, $Z \approx R_{contact\ 1} + R_{contact\ 2} + \frac{1}{\frac{1}{R_e} + i\omega(C_{geo} + C_{EDL})} \approx R_{contact\ 1} + R_{contact\ 2} + R_e$. The measured impedance then represents the electronic resistance of MGM.

In our experiment, the measured conductivity of electrolytes ranges from 33 to 107 mS cm$^{-1}$, depending on the electrolytes used. Accordingly, R$_i$ ranged from 4.4 ×10$^{-3}$ to 1.2 ×10$^{-1}$ Ω, depending on the thickness of MGM (~6 to 100 µm). The calculated C$_{EDL}$ ranges from 122.0

~ 142.3 F g$^{-1}$ (at a scan rate of 5 mV s$^{-1}$, the mass of one MGM is ~1.0 mg). Thus, the resultant frequency originating from the time constant, $\frac{1}{R_i C_{EDL}}$, is in the range from 60 to 1800 Hz, far larger than the frequency employed in the experiment (1 Hz). Besides, C$_{geo}$ is in the order of 10$^{-12}$ F cm$^{-2}$ [16] which is far smaller than the C$_{EDL}$ of MGM, the R$_e$ is ranged from 8.1×10$^{-4}$ ~ 1.3×10$^{-2}$ Ω, and the calculated $\frac{1}{R_e(C_{geo}+C_{EDL})}$ ranges from 550 to 8800 Hz. With all these considered, the impedance measured through the circuit shown in Fig. S1 at low frequency can be described as $Z \approx R_{contact\ 1} + R_{contact\ 2} + R_e$. In other words, at low frequencies (*e.g.* 1 Hz), the conducting path through ion transport was blocked due to the large capacitive reactance of the EDL. The measured impedance is mainly contributed by electronic resistance.

The validity of using AC method to obtain the electronic conductance was also confirmed by measuring the impedance of MGMs (interlayer distance: 0.6 nm) in KCl of various concentrations using conventional Electrochemical Impedance Spectroscopy. As shown Fig. S2a, the real part of the impedance (Re(Z)) of MGM-0.6 nm (freeze-dried, without electrolyte) and MGM-0.6 nm infiltrated with 1.0 M KCl overlaps at the low frequency (i.e., right side of Fig. S2a where Re(Z)=1.17 Ohm). This confirms that ionic conductance is excluded in our experimental method, as the measured impedance remains the same regardless the presence of KCl. This ionic-resistance-independent impedance is further investigated by the impedance measurement. The results show that Re(Z) at low frequency (circled in Fig. S2b) has only slight changes in 0.01 M KCl, 0.1 M and 1.0 M KCl, as opposed to the large variation of Z at remaining frequencies.

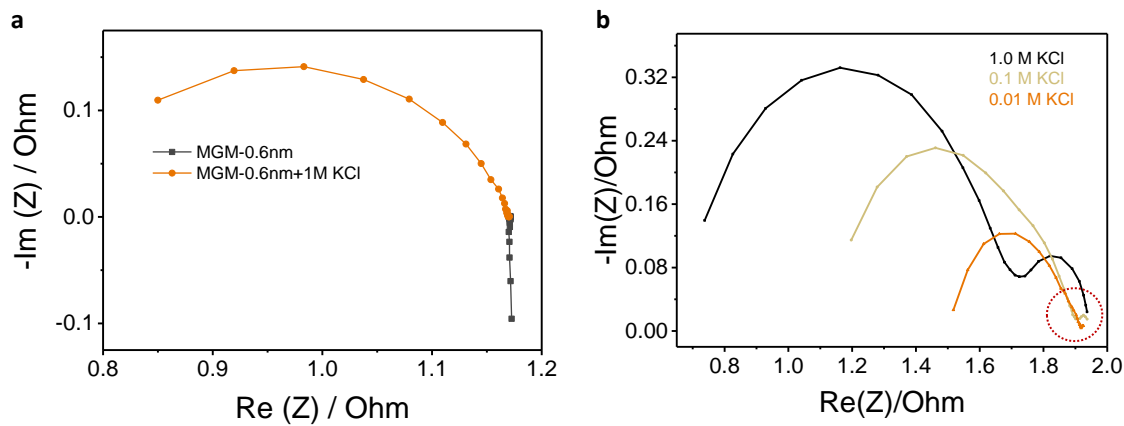

**Figure S2**. **Experimental analysis of the validity of using the AC method to measure the G$_{MGM}$. a**, Nyquist plot of MGM-0.6 nm (freeze-dried where KCl is not present, black curve) and MGM-0.6 nm infiltrated with 1.0 M KCl (orange curve). **b**, Nyquist plot of MGM-10 nm

in 0.01 (orange), 0.1 (dark yellow) and 1.0 M (black) KCl. The frequency range is 1 ~100 k Hz. The Nyquist plot of dry MGM-0.6 nm (without KCl) overlaps with that of MGM-0.6 nm infiltrated with 1.0 M KCl at low frequency, indicating that ionic conductance contributes little to the impedance at low frequencies.

As shown in Fig. S2, the contribution of ionic conductance to the impedance is negligible when the frequency is below 10 Hz. This is consistent with the discussion on Equation S1 where the impedance would be dominated by the electronic conductance for the frequency smaller than 60 Hz. To clearly show the variation of impedance with frequencies and determine the frequencies to be used in the experiments, we studied the Bode impedance of MGM-0.8 nm and MGM-5.0 nm in the testing electrolyte. As shown in Fig. S3, the log ($|Z|$) changes little when the frequency is below 10 Hz (in comparison to the frequency of 10 Hz, the variation of $|Z|$ is ~0.1 % for MGM-0.8 nm at 1 Hz, and is ~2.0 % for MGM-5.0 nm). In light of the above analysis, the frequency of 1 Hz was chosen to measure the electronic conductance of MGM.

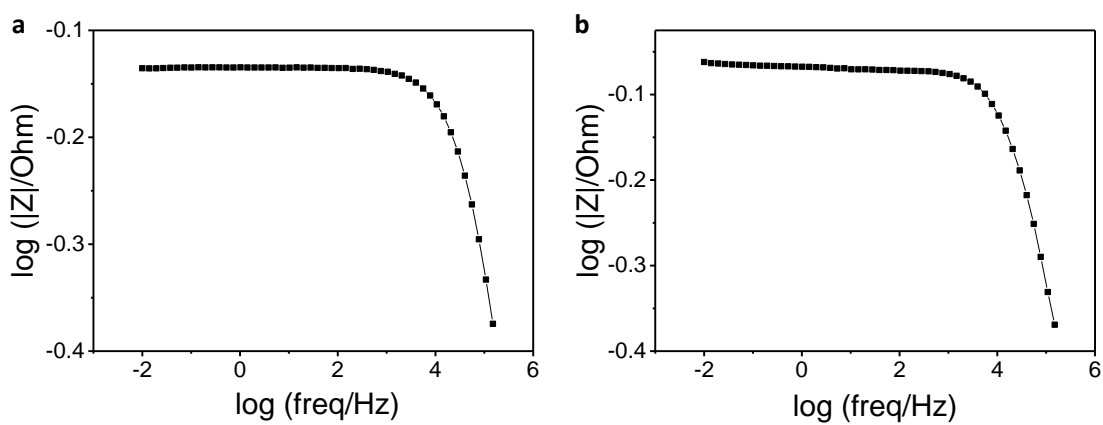

**Figure S3. Experimental determination of the frequency used for measuring $G_{MGM}$.** Bode impedance of MGM-0.8 nm (**a**) and MGM-5 nm (**b**) in 1.0 M BMIMBF$_4$. The frequency range examined is 0.01 ~150 k Hz. As shown in the figure, the log ($|Z|$) almost becomes constant when the frequency is below 10 Hz.

In summary, our equivalent circuit model and experimental analysis revealed that the decoupling of electronic and ionic contributions in MGM can be achieved through the application of an AC current with low frequencies.

**Investigation into the influence of faradaic reaction and Ohmic contact on the conductance measured**

To avoid the influence of any faradaic reaction on the electronic conductance of MGM measured, the gate potential was confined to ±0.8 V in the two-electrode device to avoid water electrolysis[3]. A CV measurement using a three-electrode system was also conducted to rule out the effect of faradaic reaction in our tests (Fig. S4). We have also conducted an I-V test to examine the nature of contact between the Pt and the MGM, which turned out to be Ohmic in the potential range we tested (Fig. S5). This result rules out the possibility that the measured conductance variation was caused by Schottky barriers[10].

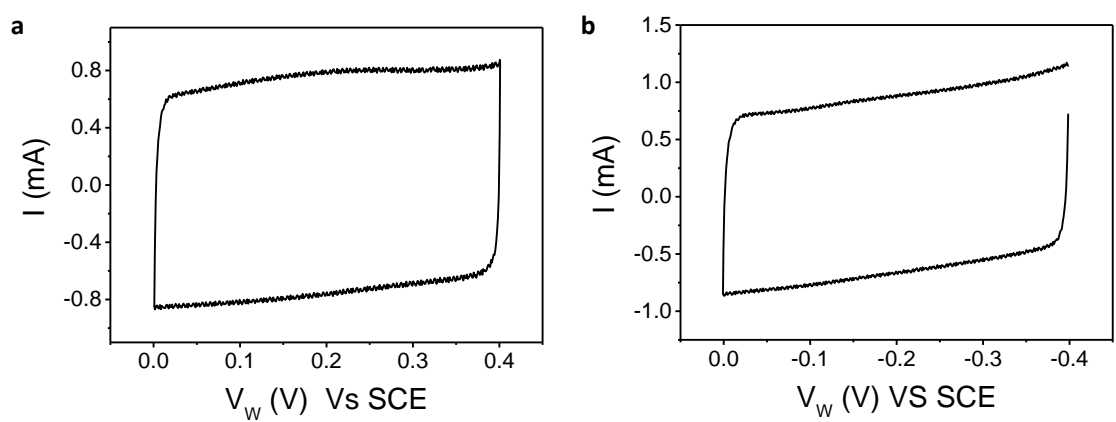

**Figure S4. Electrochemical exclusion of the influence of faradaic reactions in our experiments.** CV of MGM-10 nm in 1.0 M KCl under positive (**a**) and negative (**b**) applied potential using a three-electrode design (saturated calomel electrode, SCE, as the reference electrode). The scan rate is 5 mV s$^{-1}$. As shown in the figure, there is no obvious oxidation/reduction peak in the CV curve within ±0.4 V (vs. SCE, three electrode test), indicating that the faradaic reaction is negligible for the votlage range studied.

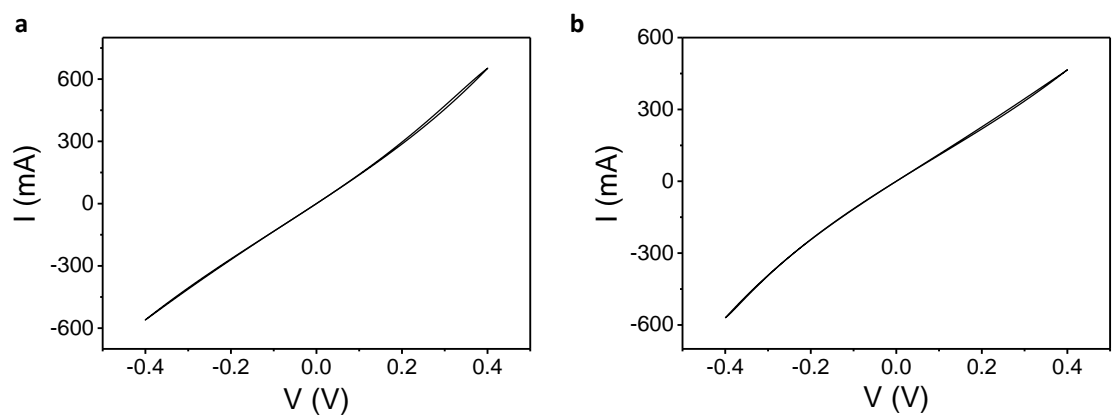

**Figure S5. Analysis of the influence of contact resistance on the G$_{MGM}$ measured.** I-V curve of MGM-0.8 nm (**a**) and MGM-5.0 nm (**b**) in 1.0 M BMIMBF$_4$. The voltage was applied between the Pt ring and the Pt current collector. As shown in the figure, the I-V curve is linear

in the range of -0.2 V to 0.2 V, indicating that the contact between the MGMs and the Pt is Ohmic at a potential of 10 mV we used and it contributes little to the variation of $G_{MGM}$ during the electrolyte gating process.

**Examining the nature of electrolyte gating effect**

Given that the change of $G_{MGM}$ with gate voltages is mainly a result of the electrolyte gating effect, the electrical response of MGM to gate voltages should follow a similar trend to that of rGO which is the basic building blocks of MGM. To further confirm the nature of electrolyte gating, we have compared our results with the electrical gating properties of single rGO sheets based on the traditional back-gating device configuration as shown in Fig. 2a.

The conductivity of MGM ($\sigma_{MGM}$, a geometry-independent specific conductance) can be written in terms of electron/hole mobility ($\mu$) and charge carrier density ($n_{ind}$)[9],

$$\sigma_{MGM}(V_g) = \mu e n_{ind}(V_g) \tag{S2}$$

where $V_g$ is the gate voltage potential applied between the gate electrode and MGM. In the electrical studies of monolayer/few-layer graphene, both simple conductivity and Hall effect measurements indicate that charge carrier density ($n_{ind}$) in graphene could be estimated by considering the mobility of electron/hole ($\mu$) as a gate-voltage-independent constant[17]. In graphene-electrolyte systems, an electrolyte-gated Hall effect measurement confirms that the change of $\mu$ with gate voltage ($V_g$) could also be negligible when $V_g$ is away from the neutrality point (0.1 V away)[18,19]. Therefore, in this work, the variation in conductance of MGM ($G_{MGM}$) can be considered as a result of the charge carrier density in MGM when $V_g$ is kept away from the neutrality point (under negative gating, 0.4 V away in Fig. 2b)[20].

In conventional back-gating setups using monolayer graphene, charge carriers in graphene are induced from a capacitive charging process through a traditional capacitor (with the capacitance $C_{BG}$) underneath graphene. The induced density is a function of $V_g$,

$$en_{ind} = C_{BG}(V_g - V_{BG}^0) \tag{S3}$$

where $V_{BG}^0$ is the previously mentioned neutrality point for back-gating.

Similarly, in the electrolyte-gating setup, charge carriers in graphene are induced from a capacitive charging process through a capacitor composed of electrical double layer (EDL) formed near graphene-electrolyte interfaces. A similar relation to eq. S3 is written as

$$en_{ind} \approx C_{EDL}(V_g - V_{EDL}^0) \tag{S4}$$

where $C_{EDL}$ is the capacitance of EDL and $V_{EDL}^0$ is the neutrality point for electrolyte-gating.

Note that although, strictly speaking, $en_{ind} = C_{EDL}\phi_{EDL}$, where $\phi_{EDL}$ is the potential drop only across the EDL. Eq. S4 still provides a reasonable approximation for our study of MGM under electrolyte-gating. Since $V_g$ is the voltage measured across the entire graphene-electrolyte interface (including both $\phi_{EDL}$ and the potential drop inside monolayer graphene ($\delta$)), $\phi_{EDL}$, can be written as $\phi_{EDL} = V_g - \delta$. Note that $\delta$ is proportional to the ratio of $C_{EDL}$ and the quantum capacitance of graphene ($C_Q$, an intrinsic property), and $C_Q$ is much greater than $C_{EDL}$ in electrolyte-gating setups when $V_g$ is away from the neutrality point[21], we obtain $\phi_{EDL} \approx V_g$ and hence eq. S4.

In comparison to the $G_{MGM}$ which is dependent on the geometry of the MGM, such as the number of rGO sheets composing the MGM, the $n_{ind}$ is rather independent of the geometry of MGM. Then, we apply the above relations to compare the changes in $n_{ind}$ of MGM under electrolyte-gating with that of rGO under conventional back-gating (the latter is commonly seen in the electrical studies of graphene, *e.g.*, electronic transistors) to further prove the electrolyte gating effect in MGM.

In order to examine the variation of charge carriers in rGO as a function of gate voltage, a back-gated device is fabricated using a monolayer reduced graphene oxide (rGO) (inset in Fig. 2a). Briefly, the aforementioned rGO solution (0.5 mg g$^{-1}$) was diluted to 0.005 mg g$^{-1}$ and then drop-casted onto a marked silicon oxide (285 nm). After being dried at room temperature overnight, a pair of electrodes (source-drain electrode with a distance of 5 μm) were then patterned using direct UV-laser writing lithography (SF-100) followed by electron-beam deposition of Ti/Au (5/50 nm) in the vacuum. Afterwards, the devices were annealed at 200 ℃ for 2 h in argon atmosphere prior to testing. The source-drain current was collected using a two-channel source meter unit (Agilent, B2902A) in ambient condition. While $C_{BG}$ of this particular device is 1.26×10$^{-8}$ F cm$^{-2}$, $C_{EDL}$ is estimated to be 142.3 F g$^{-1}$ from CV measurements. These numbers are then substituted in eqs. S3 and S4 for the calculation of charge carrier density ($n_{ind}$).

It is worth noting that the exact surface area of the MGM needs to be precisely measured to obtain the density of charge carriers, which, however, is difficult[3]. Previous studies reported that the specific surface area of graphene assemblies can range from 961 [3] to 3100 m$^2$ g$^{-1}$ [2] showing a dependence on the preparation and calculation methods. Since the rGO sheets in the MGM remain mostly separated[2,3], a rough estimation can be made to estimate the carrier density of MGM by assuming the specific surface area to be 2630 m$^2$ g$^{-1}$ (theoretical value of graphene). As shown in Fig. 2, when the charge carrier density for MGM and rGO (shown on top x-axes) increases from 0 to around 90×10$^{11}$ cm$^{-2}$, G$_{MGM}$ rises by 26%, which is close to the change of G$_{rGO}$ (18%). Note, though, the value of $n_{ind}$ in Fig. 2b is only indicative, as the exact surface area of MGM which contributes to electrical conductance is difficult to estimate. The consistence in the variation of conductance with gate voltage, as well as the charge carrier density, indicates that the change of G$_{MGM}$ is mainly a result of the electrolyte gating effect.

## 2. Supplementary Figures (Fig. S6- S15)

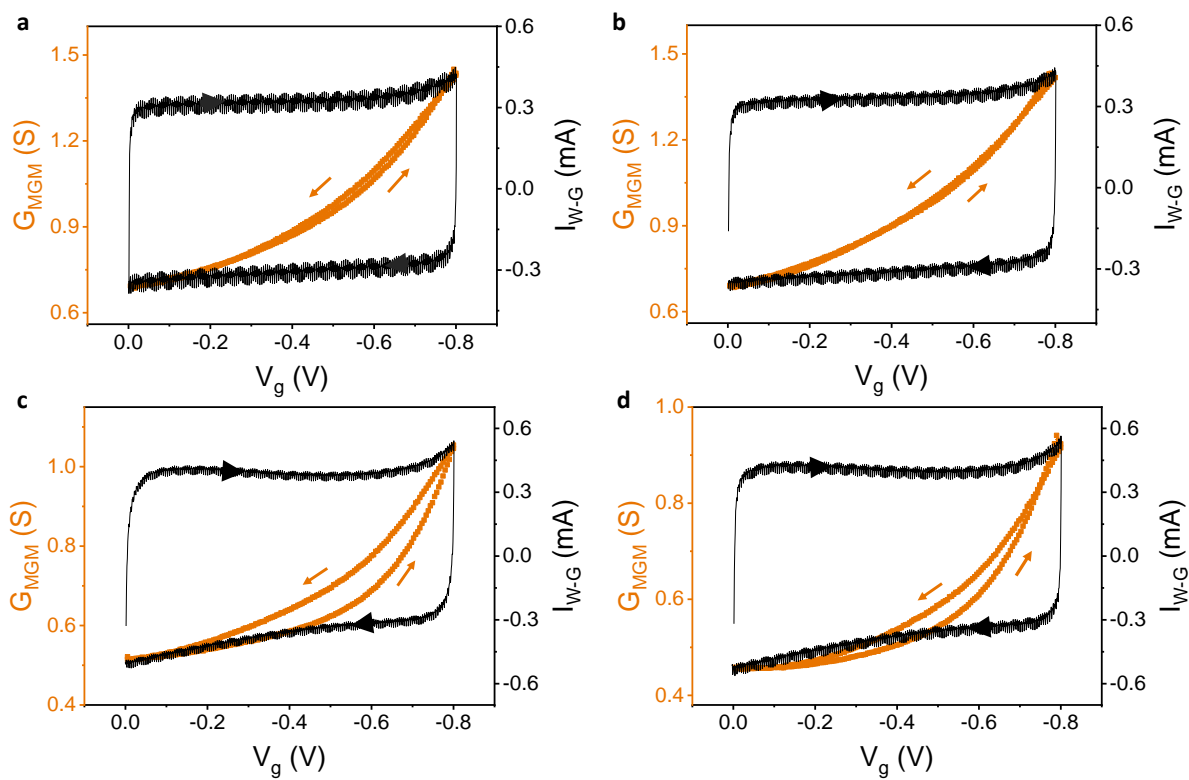

**Figure S6. The effect of electrolytes on the response of $G_{MGM}$ (MGM-10 nm) to $V_g$ in different electrolytes. a-d**, $G_{MGM}$ and $I_{W-G}$ as a function of $V_g$ in 1.0 M KCl (**a**), 0.5 M KTFSI (**b**), 1.0 M EMIMBF$_4$ (**c**) and 1.0 M BMIMBF$_4$ (**d**). The scan rate is 5 mV s$^{-1}$. For KTFSI, the concentration of 0.5 M was used due to its limit of solubility. As shown in the figure, $G_{MGM}$ consistently shows ~100% increase from 0 to -0.8 V in electrolytes of different cations or anions.

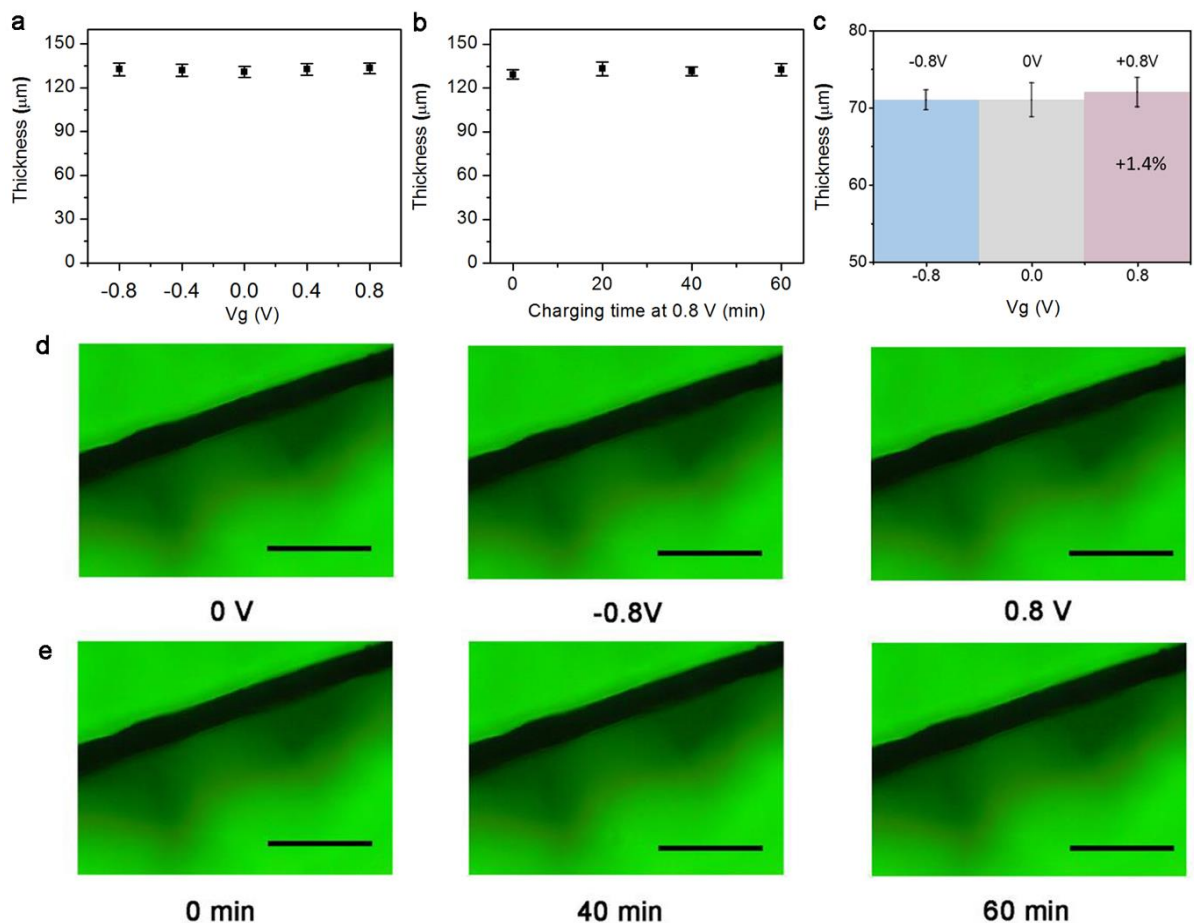

**Figure S7.** *In situ* **optical measurement of the thickness of an MGM-5.0 nm upon charging.** a) The membrane thickness upon charging with different voltage amplitude. The charging time is 5 mins. b) The membrane thickness upon 0.8 V charging with different charging time. c) Micrometre measurement of membranes thickness upon charging with different voltage amplitude. The charging time is 5 mins. d) Typical optical images of the cross section of graphene membranes under 0, -0.8 V and 0.8 V. e) Typical optical images of the cross section of graphene membranes under 0.8 V at 0 min, 40 mins and 60 mins. The scale bar in d) and e) is 500 μm.

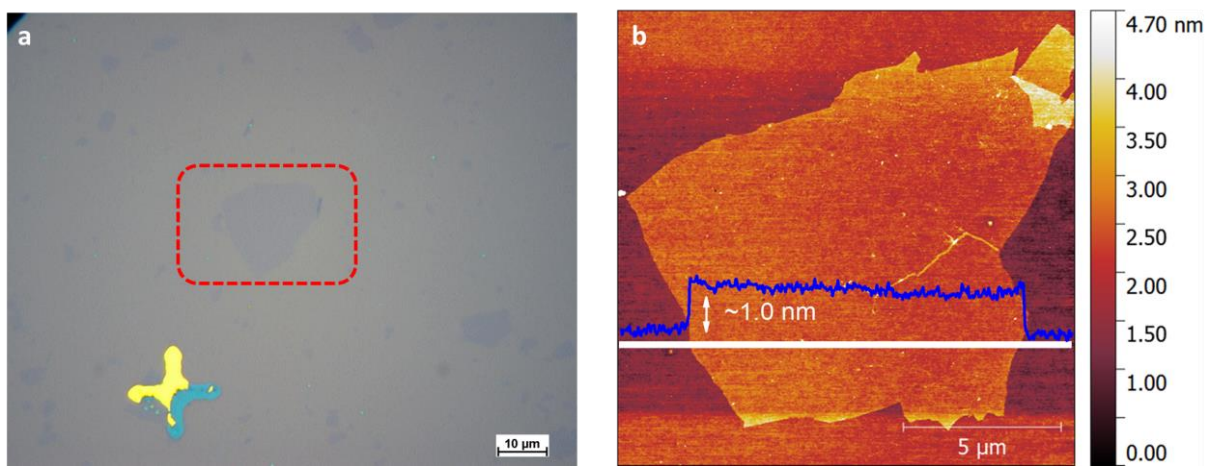

**Figure S8. Characterization of the rGO used for back gating.** A microscope image of the rGO on silicon oxide (**a**), An AFM image of the rGO sheet (**b**). The measured thickness of the rGO is ~ 1.0 nm, indicating that rGO sheets in dispersion remained separated before vacuum filtration.

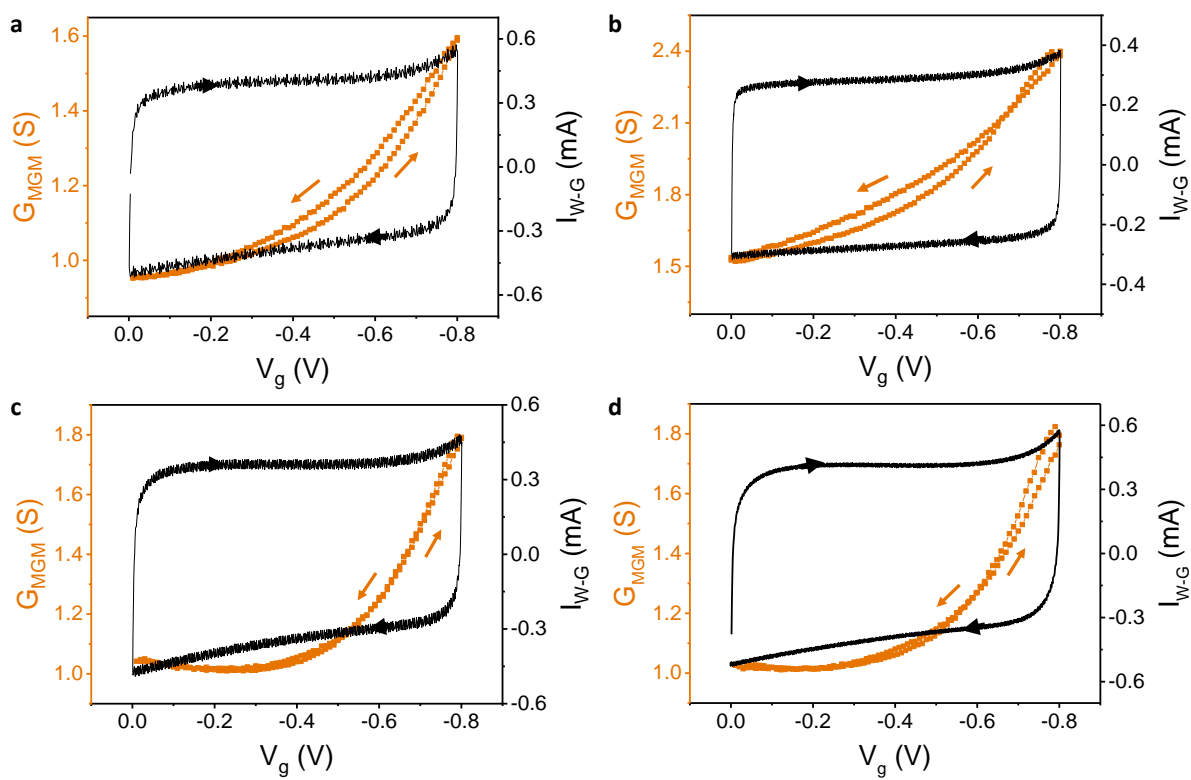

**Figure S9. Effect of the type of electrolytes on the response of $G_{MGM}$ (MGM-5.0 nm) to $V_g$. a-d**, $G_{MGM}$ and $I_{W-G}$ as a function of $V_g$ in 1.0 M KCl (**a**), 0.5 M KTFSI (**b**), 1.0 M EMIMBF$_4$ (**c**) and 1.0 M BMIMBF$_4$ (**d**). The scan rate is 5 mV s$^{-1}$. As shown in the figure, $G_{MGM}$ shows a consistent variation with $V_g$ in various electrolytes, indicating that the response

of $G_{MGM}$ to $V_g$ in nanoporous MGM is not ion specific. In comparison to the response of $G_{MGM}$ to $V_g$ for MGM-10 nm, which shows ~100% increase at -0.8 V, $G_{MGM}$ of MGM-5.0 nm shows a smaller increase (~ 60%), which can be attributed to the higher stacking level of rGO sheets in MGM-5.0 nm[3,22].

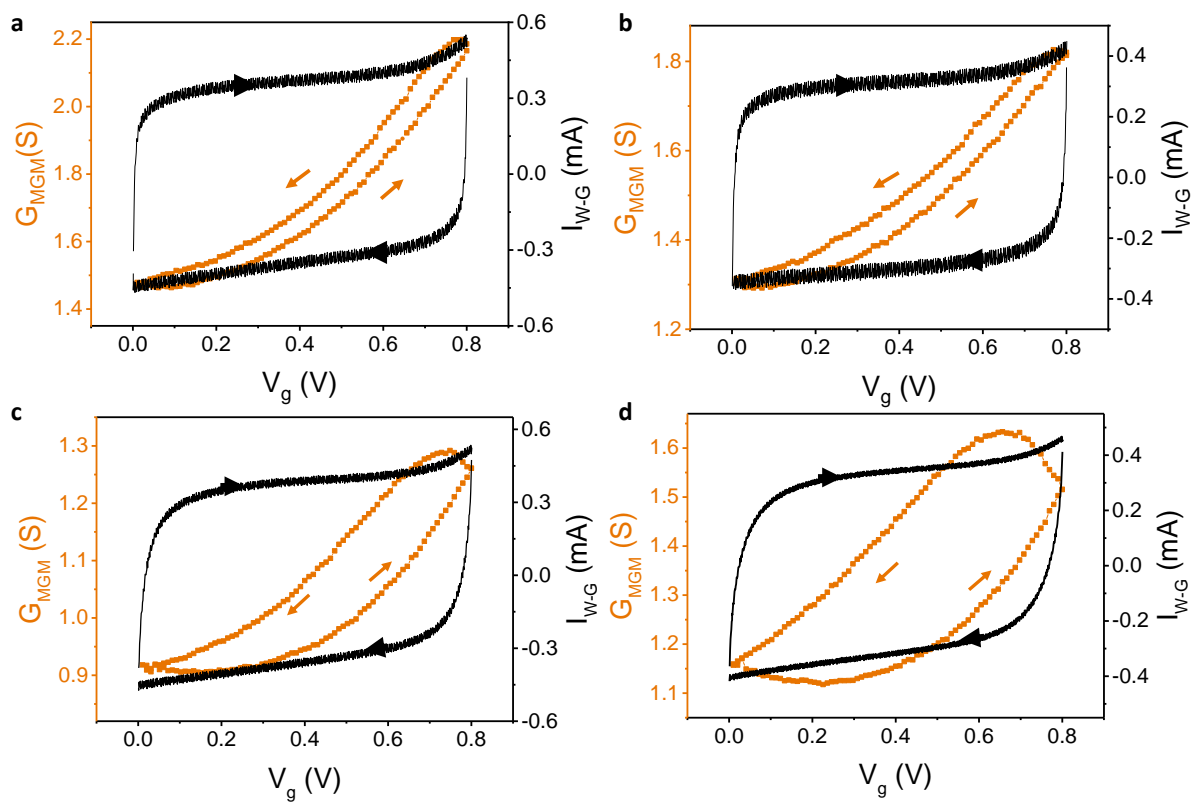

**Figure S10. Effect of type of electrolytes on the hysteretic response of $G_{MGM}$ (MGM-0.8 nm) to $V_g$. a-d**, $G_{MGM}$ and $I_{W-G}$ as a function of $V_g$ in 1.0 M KCl (**a**), 0.5 M KTFSI (**b**), 1.0 M EMIMBF$_4$ (**c**) and 1.0 M BMIMBF$_4$ (**d**). The scan rate is 5 mV s$^{-1}$. An obvious hysteretic response of $G_{MGM}$ with regard to $V_g$ can be also found in 0.5 M KTFSI and 1.0 M EMIMBF$_4$, in addition to 1.0 M BMIMBF$_4$, implying that the observed hysteresis is not ion specific. The obvious decrease in the hysteresis from BMIMBF$_4$ to EMIMBF$_4$, KTFSI and finally KCl implies that the hysteresis is likely correlated to the ionic size of the electrolyte used, i.e., the spatial confinement of ions. Besides, in comparison to the rise of $G_{MGM}$ for MGM-10 nm (~100%) and MGM-5.0 nm (~60%) when $V_g$ is varying from 0 to -0.8 V, the increase of $G_{MGM}$ for MGM-0.8 nm (~40%) further decreases, which can be explained by the additional increase in the stacking level of rGO sheets within MGM-0.8 nm[3,22].

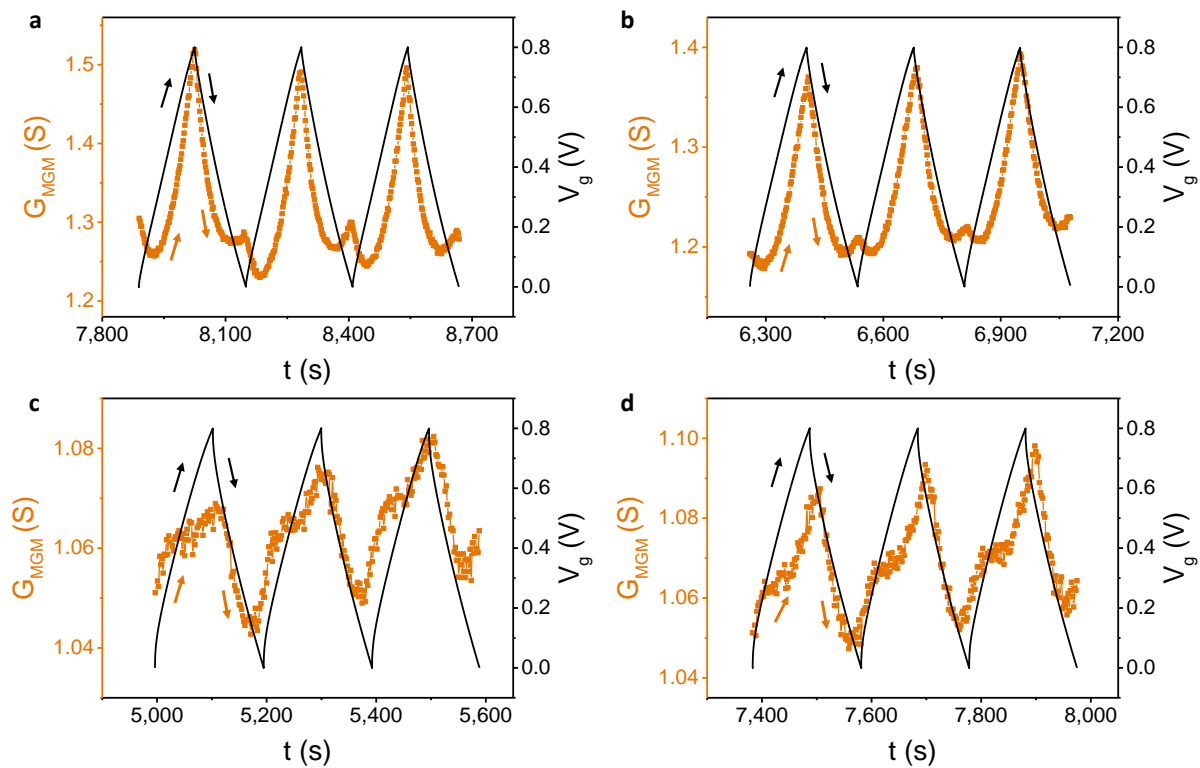

**Figure S11. Characterization of the hysteretic response of $G_{MGM}$ under positive polarisation. a-d**, $G_{MGM}$ and $V_g$ as a function of time for MGM with an interlayer spacing of 5.0 nm (**a**), 2.0 nm (**b**), 0.8 nm (**c**) and 0.6 nm (**d**). The electrolyte is 1.0 M BMIMBF$_4$ and the charging rate is 0.5 A g$^{-1}$. As shown in the figure, though $G_{MGM}$ varies in a non-linear way under positive polarisation, we can still read the hysteresis from the time lag between the peak of $G_{MGM}$ and the peak of $V_g$, which is distinct in MGM with sub-nanometre interlayer distance (Fig. S11c and d).

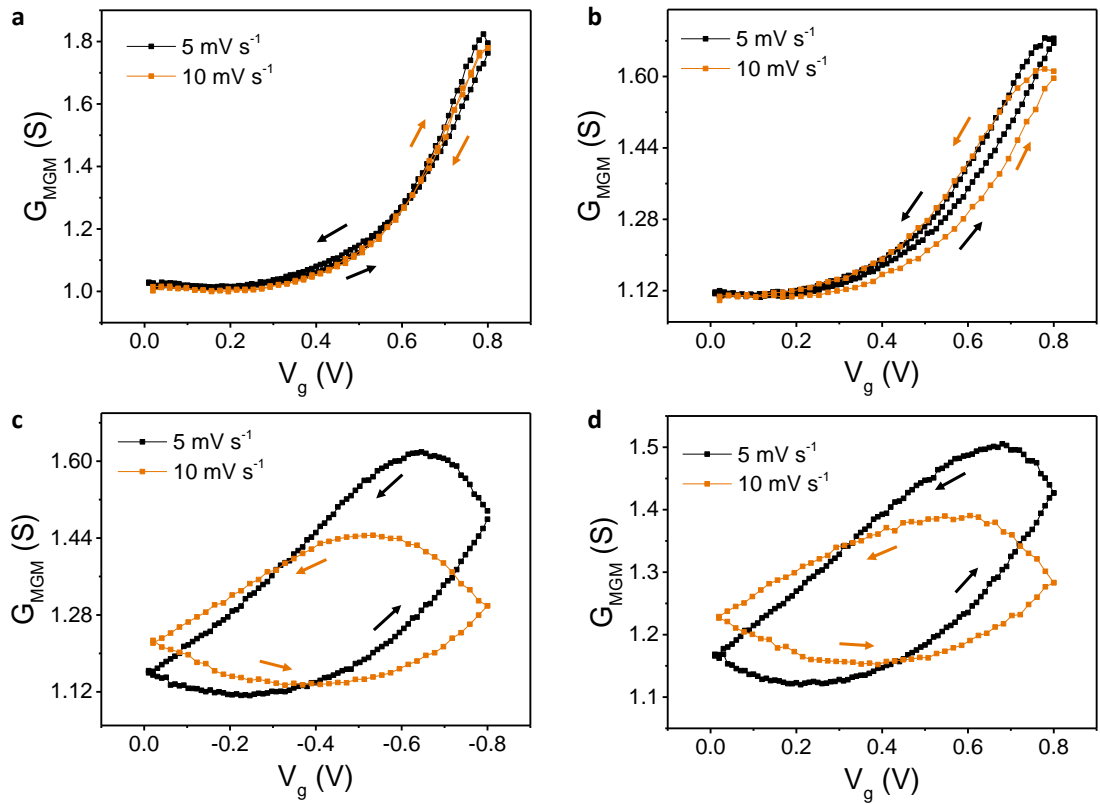

**Figure S12. Characterization of the hysteretic response of $G_{MGM}$ at various scanning rates. a-d**, $G_{MGM}$ as a function of $V_g$ for MGM with an interlayer spacing of 5.0 nm (**a**), 2.0 nm (**b**), 0.8 nm (**c**) and 0.6 nm (**d**), respectively. The electrolyte used is 1.0 M BMIMBF$_4$. As shown in the figure, for MGM with sub-nanometre interlayer distance, the hysteresis loop becomes more prominent at higher scanning rates, indicating a stronger hysteresis[23,24].

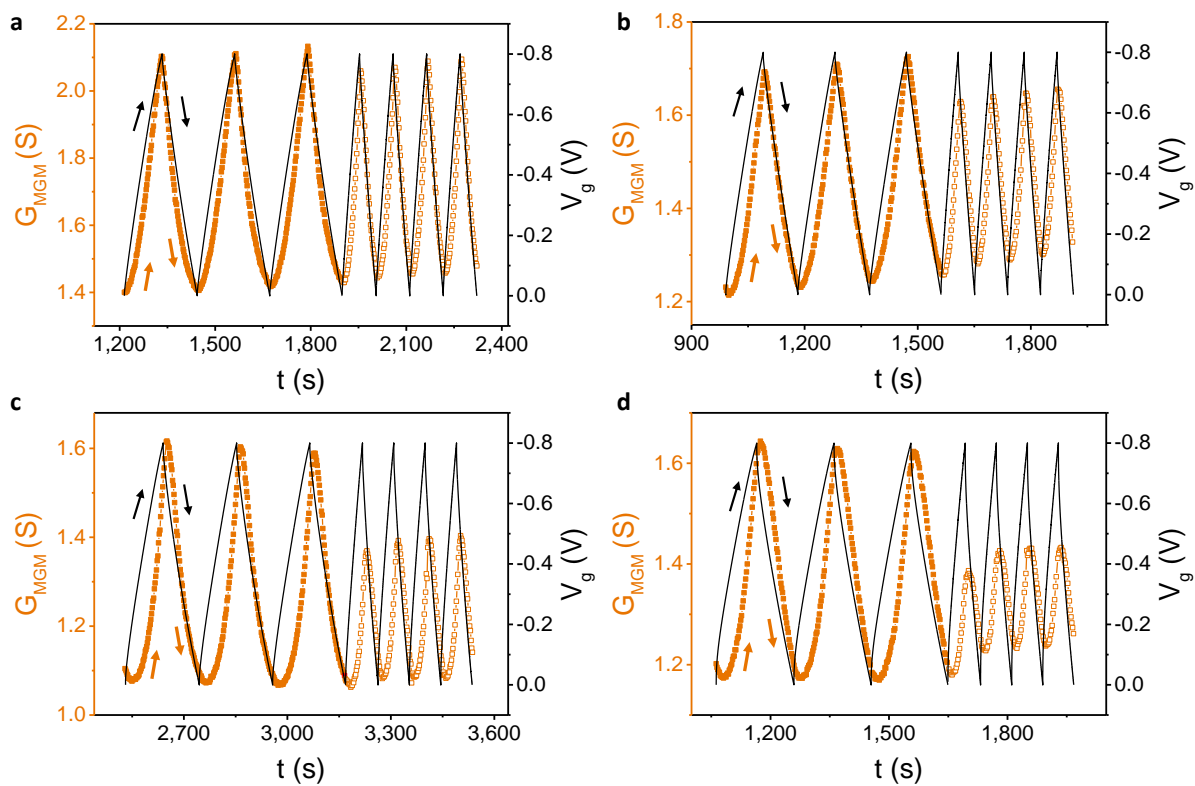

**Figure S13. The effect of electrolytes on the change of the lowest value of $G_{MGM}$ with charging rates. a-d**, $G_{MGM}$ and $V_g$ as a function of time in 1.0 M KCl (**a**), 0.5 M KTFSI (**b**), 1.0 M EMIMBF$_4$ (**c**) and 1.0 M BMIMBF$_4$ (**d**). The interlayer spacing of the MGM is 0.8 nm. The orange curve with solid square denotes a charging rate of 0.5 A g$^{-1}$ while the curve with open square signifies a charging rate of 1.0 A g$^{-1}$. As shown in the figure, the lowest value of $G_{MGM}$ increases with charging rates in all four electrolytes examined, and the most distinct increase occurs in 1.0 M BMIMBF$_4$. Thus, this increase in the minimum value of $G_{MGM}$ with charging rates seems to be generic in MGM with small nanopores.

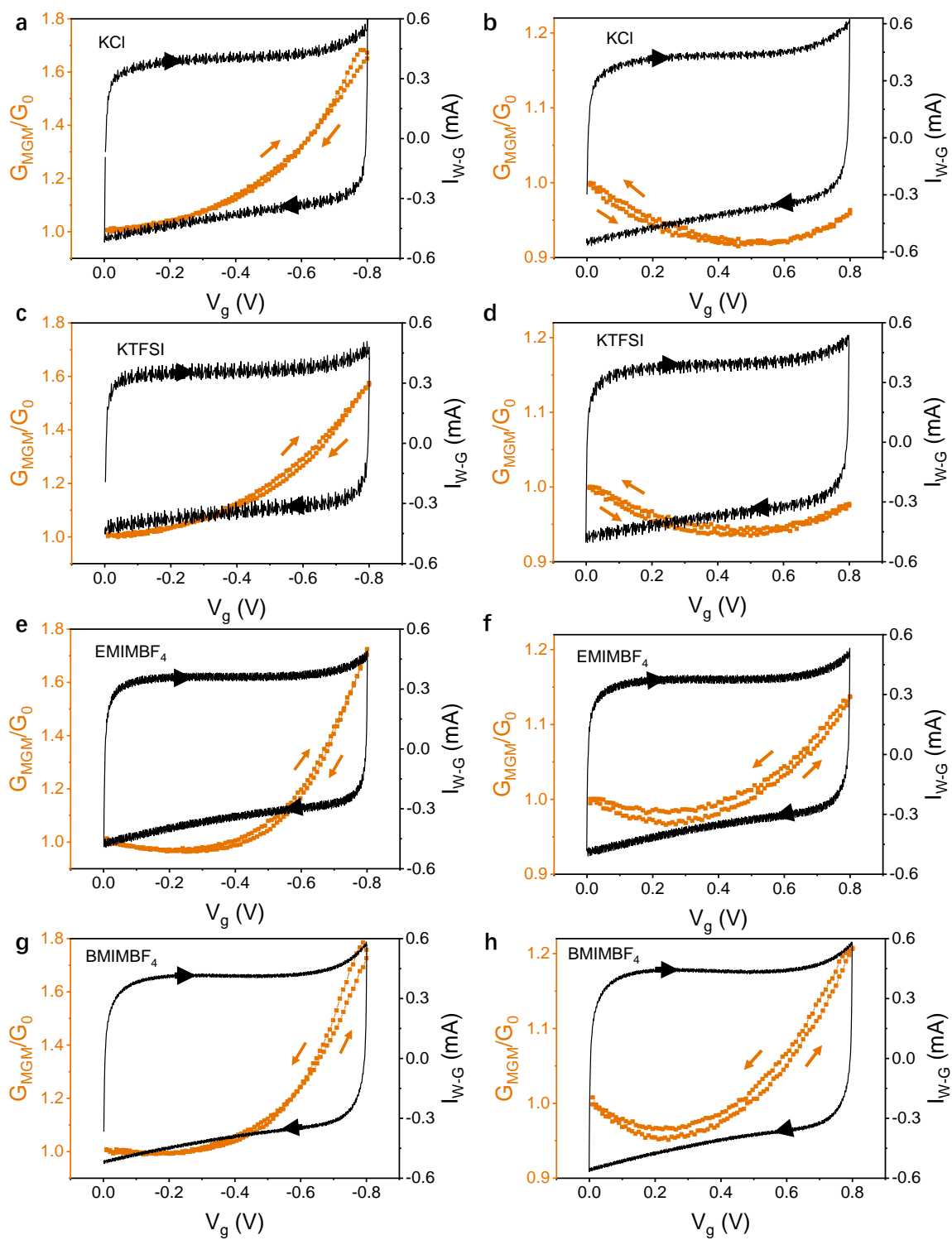

**Fig. S14. Comparison of the electrolyte-gating effect of the MGM-5.0 nm sample in different electrolytes under negative (left column) and positive (right column) polarisations.** The aqueous solution of electrolytes used in the measurements are (**a,b**), 1.0 M KCl, (**c,d**), 0.5 M KTFSI, (**e,f**), 1.0 M EMIMBF$_4$, and (**g,h**) 1.0 M BMIMBF$_4$. Note that KCl and KTFSI share the same cation while EMIMBF$_4$ and BMIMBF$_4$ contain the same anion. G$_0$ represents the value of G$_{MGM}$ located at V$_g$=0.

As can be seen from Fig. S14, the $G_{MGM}$-$V_g$ relationship under negative polarisation is mostly monotonic, while $G_{MGM}$ under positive polarisation increases at a certain voltage after reaching a minimum value. As discussed in the main text, the occurrence of minimum of $G_{MGM}$ during positive polarisation can be ascribed to the shifted "Dirac" point due to the chemical doping of graphene sheets in MGM. This chemical doping may be contributed by the functional groups or surface charges present in MGM and the electrostatic interaction between graphene and the electrolyte described by the Gerischer model[26]. As a result, the $G_{MGM}$-$V_g$ relation in negative polarisation is monotonic, while it is non-monotonic under positive polarisation. As the $G_{MGM}$-$V_g$ relation in negative polarisation is away from the "Dirac" point, the number of majority charge carriers in graphene should be proportional to the number of ions in EDL. Therefore, the monotonic relation suggests that the number of charge carriers in graphene changes, and hence the number of ions in EDL, with the applied voltage. This indicates that the monotonic $G_{MGM}$-$V_g$ relation under negative polarisation could be used to study the charging/discharging process in EDL. The non-monotonic $G_{MGM}$-$V_g$ relation under positive polarisation, on the other hand, explicitly includes not only the information of ion transport in EDL but also their interaction with rGO. It can be seen from Figs. S14 b, d, f, and h that the values of $V_g$ where $G_{MGM}$ reaches its minimum are different when the MGM is placed in different electrolytes. This observation is consistent with our understanding that the interaction between charge carriers in rGO and electrolyte would contribute to the shift of the minimum of $G_{MGM}$, and hence the non-monotonic $G_{MGM}$-$V_g$ relation. The CV curves shown in Figs. S14 b, d, f, and h clearly demonstrate the charging and discharging processes in forward and backward scans, where the number of net charges (cations subtracted by anions) obtained by integration of current over time monotonically increases and decreases, respectively. However, the exact numbers of cations and anions are unknown. For example, the net charge of electrolyte confined in MGM could still be 0 at $V_g$=0 due to electroneutrality. It is therefore unclear which species of ions has larger effects on the non-monotonic $G_{MGM}$-$V_g$ relation under positive polarisation. However, it is reasonable to postulate that the non-monotonic $G_{MGM}$-$V_g$ relation demonstrate a desorption of anions followed by adsorption of cations on graphene-electrolyte interfaces in a forward scan in the right column of Fig. S14, as a depletion of holes and accumulation of electrons in graphene could be observed.

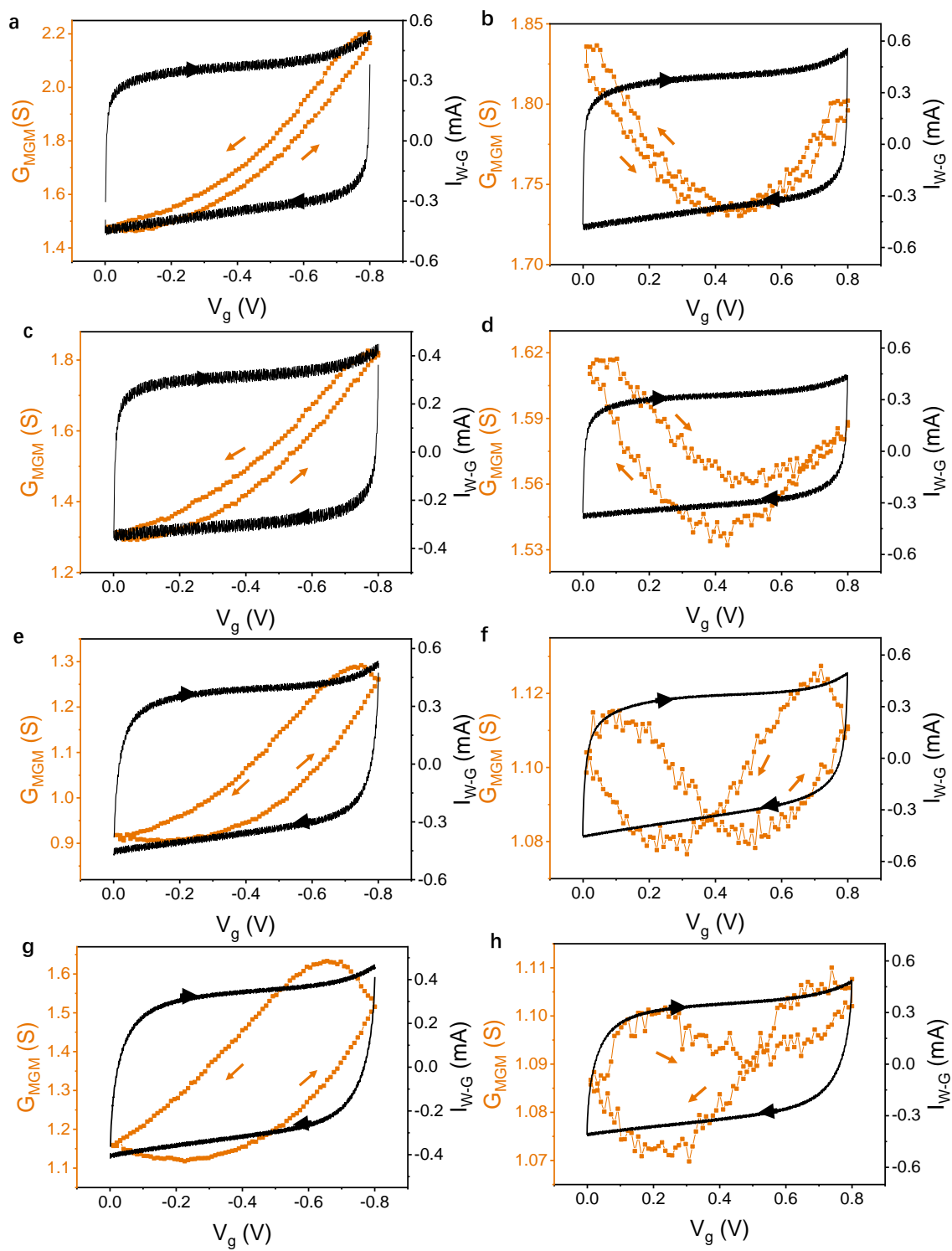

**Fig. S15. Comparison of the electrolyte-gating effect of the MGM-0.8 nm sample in different electrolytes under negative (left column) and positive (right column) polarisations.** The aqueous solution of electrolytes used in the measurements are (**a,b**), 1.0 M KCl, (**c,d**), 0.5 M KTFSI, (**e,f**), 1.0 M EMIMBF$_4$, and (**g,h**) 1.0 M BMIMBF$_4$. As the interlayer spacing becomes smaller (0.8 nm in comparison with 5.0 nm in Fig. S14), the hysteresis effect in $G_{MGM}$ discussed in the main text

can be observed in both negative and positive polarizations. Again, the shape of the hysteresis is varied with the type of electrolytes tested. As discussed in the main text, the charging dynamics of nanopores-based supercapacitors is very complex and depends on a broad range of parameters such as the electrode and electrolyte materials used, the size and surface chemistry of pores, the polarization of the electrode, the operation rate, as well as the ion concentration, the cation-to-anion size ratio, solvation energies, ionophobicity, and ion-ion correlations. In general, these results appear to be in line with the previous findings obtained from other techniques.

## 3. Supplementary References:

1. Li, D., Mueller, M. B., Gilje, S., Kaner, R. B. & Wallace, G. G. Processable aqueous dispersions of graphene nanosheets. *Nat. Nanotechnol.* **3**, 101-105 (2008).
2. Yang, X. *et al.* Ordered gelation of chemically converted graphene for next‐generation electroconductive hydrogel films. *Angew. Chem. Int. Edit.* **50**, 7325-7328 (2011).
3. Yang, X., Cheng, C., Wang, Y., Qiu, L. & Li, D. Liquid-mediated dense integration of graphene materials for compact capacitive energy storage. *Science* **341**, 534-537 (2013).
4. Cheng, C., Jiang, G., Simon, G. P., Liu, J. Z. & Li, D. Low-voltage electrostatic modulation of ion diffusion through layered graphene-based nanoporous membranes. *Nat. Nanotechnol.* (2018).
5. Cheng, C. *et al.* Ion transport in complex layered graphene-based membranes with tuneable interlayer spacing. *Sci. Adv.* **2**, e1501272 (2016).
6. Eda, G., Fanchini, G. & Chhowalla, M. Large-area ultrathin films of reduced graphene oxide as a transparent and flexible electronic material. *Nat. Nanotechnol.* **3**, 270-274 (2008).
7. He, Q. *et al.* Transparent, flexible, all-reduced graphene oxide thin film transistors. *ACS Nano* **5**, 5038-5044 (2011).
8. Wang, S. *et al.* High mobility, printable, and solution-processed graphene electronics. *Nano Lett.* **10**, 92-98 (2009).
9. Das, A. *et al.* Monitoring dopants by Raman scattering in an electrochemically top-gated graphene transistor. *Nat. Nanotechnol.* **3**, 210-215 (2008).
10. Heller, I. *et al.* Influence of electrolyte composition on liquid-gated carbon nanotube and graphene transistors. *J. Am. Chem. Soc.* **132**, 17149-17156 (2010).
11. Ohno, Y., Maehashi, K., Yamashiro, Y. & Matsumoto, K. Electrolyte-gated graphene field-effect transistors for detecting pH and protein adsorption. *Nano Lett.* **9**, 3318-3322 (2009).
12. Albery, W. J., Elliott, C. M. & Mount, A. R. A transmission line model for modified electrodes and thin layer cells. *J. Electroanal. Chem.* **288**, 15-34 (1990).
13. Jamnik, J. & Maier, J. Treatment of the impedance of mixed conductors equivalent circuit model and explicit approximate solutions. *J. Electrochem. Soc.* **146**, 4183-4188 (1999).
14. Riess, I. Mixed ionic–electronic conductors—material properties and applications. *Solid State Ion.* **157**, 1-17 (2003).
15. Saab, A. P., Garzon, F. H. & Zawodzinski, T. A. Determination of ionic and electronic resistivities in carbon/polyelectrolyte fuel-cell composite electrodes. *J. Electrochem. Soc.* **149**, A1541-A1546 (2002).
16. Thangadurai, V., Huggins, R. A. & Weppner, W. Use of simple ac technique to determine the ionic and electronic conductivities in pure and Fe-substituted $SrSnO_3$ perovskites. *J. Power Sources* **108**, 64-69 (2002).
17. Novoselov, K. S. *et al.* Electric field effect in atomically thin carbon films. *Science* **306**, 666-669 (2004).
18. Zhan, H., Cervenka, J., Prawer, S. & Garrett, D. J. Molecular detection by liquid gated Hall effect measurements of graphene. *Nanoscale* **10**, 930-935 (2018).
19. Kim, H., Kim, B. J., Sun, Q., Kang, M. S. & Cho, J. H. Graphene Transistors Gated by Salted Proton Conductor. *Adv. Electron. Mater.* **2**, 1600122 (2016).
20. Kim, C.-H. & Frisbie, C. D. Determination of Quantum Capacitance and Band Filling Potential in Graphene Transistors with Dual Electrochemical and Field-Effect Gates. *J. Phys. Chem. C* **118**, 21160-21169 (2014).
21. Zhu, Y. *et al.* Carbon-based supercapacitors produced by activation of graphene. *Science* **332**, 1537-1541 (2011).
22. Partoens, B. & Peeters, F. From graphene to graphite: Electronic structure around the K point. *J. Phys. Rev. B* **74**, 075404 (2006).
23. Krems, M., Pershin, Y. V. & Di Ventra, M. Ionic memcapacitive effects in nanopores. *Nano Lett.* **10**, 2674-2678 (2010).

24  Wang, D. *et al.* Transmembrane potential across single conical nanopores and resulting memristive and memcapacitive ion transport. *J. Am. Chem. Soc.* **134**, 3651-3654 (2012).
25  Weppner, W. & Huggins, R. A. Determination of Kinetic-Parameters of Mixed-Conducting Electrodes and Application to System Li3sb. *J. Electrochem. Soc.* **124**, 1569-1578 (1977).
26  Memming, R. Semiconductor Electrochemistry. *Weinheim, Germany: WILEY-VCH* (2015).